\documentclass[11pt, a4paper]{article}
\pdfoutput=1

\usepackage{anysize}
\marginsize{3cm}{3cm}{1.5cm}{1.5cm}
\usepackage[T1]{fontenc}
\usepackage{inputenc}

\usepackage{amsfonts}
\usepackage{amssymb}
\usepackage{graphics}
\usepackage{epsfig}
\usepackage{graphicx}
\usepackage{epsfig}
\usepackage{amssymb}
\usepackage{amsfonts}

\usepackage{amsmath}
\usepackage{xcolor}
\usepackage{float}
\usepackage{enumerate}
\usepackage{slashed}
\usepackage{hyperref}

\hypersetup{colorlinks=true,linkcolor=blue,citecolor=blue} 
\usepackage[numbers,sort&compress]{natbib}

\usepackage{caption}
\usepackage{subcaption}

\usepackage{tikz}

\numberwithin{equation}{section}

\newcommand {\beq} {\begin{equation}}
\newcommand {\eeq} {\end{equation}}
\newcommand{\bea}{\begin{eqnarray}}
\newcommand{\eea}{\end{eqnarray}}
\newcommand{\bit}{\begin{itemize}}
\newcommand{\eit}{\end{itemize}}
\def\nl{\nonumber \\}
\def\Tr{{\rm Tr}}

\def\a{\alpha}

\def\b{\beta}

\def\s{\sigma}

\def\ep{\epsilon}

\def\p{\partial}

\def\le{\left(}
\def\ri{\right)}

\def\beq{\begin{equation}}
\def\eeq{\end{equation}}

\def\vphi{{\varphi}}

\def\hhz{{\hat{z}}}
\def\hhx{{\hat{x}}}
\def\hht{{\hat{t}}}
\def\hhphi{{\hat{\varphi}}}
\def\ta{{\tilde{\alpha}}}
\def\tb{{\tilde{\beta}}}
\def\d{{\delta}}

\begin{document}

\begin{titlepage}

\begin{flushright}

\end{flushright}
\bigskip
\begin{center}
{\LARGE  {\bf
A falling magnetic monopole \\ as a holographic local quench
  \\[2mm] } }
\end{center}
\bigskip
\begin{center}
  {\large \bf Nicol\`o Zenoni$^{a,b,c}$},
{\large \bf  Roberto  Auzzi$^{a,b}$},
 {\large \bf Stefania Caggioli$^{a}$}, \\
  {\large \bf Maria Martinelli$^{a}$}  {\large \bf and }   
   {\large \bf Giuseppe Nardelli$^{a,d}$}
\vskip 0.20cm
\end{center}
\vskip 0.20cm 
\begin{center}
$^a${ \it \small  Dipartimento di Matematica e Fisica,  Universit\`a Cattolica
del Sacro Cuore, \\
Via Musei 41, 25121 Brescia, Italy}
\\ \vskip 0.20cm 
$^b${ \it \small{INFN Sezione di Perugia,  Via A. Pascoli, 06123 Perugia, Italy}}
\\ \vskip 0.20cm 
$^c${ \it \small{Instituut voor Theoretische Fysica, KU Leuven, Celestijnenlaan 200D, B-3001 Leuven, Belgium }}
\\ \vskip 0.20cm 
$^d${ \it \small{TIFPA - INFN, c/o Dipartimento di Fisica, Universit\`a di Trento, \\ 38123 Povo (TN), Italy} }
\\ \vskip 0.20cm 
E-mails: nicolo.zenoni@unicatt.it, roberto.auzzi@unicatt.it, \\ stefy.caggioli@gmail.com, maria.martinelli1996@gmail.com, \\
giuseppe.nardelli@unicatt.it
 \end{center}
\vspace{3mm}

\begin{abstract}
An analytic static monopole solution is found in global AdS$_4$,
in the limit of small backreaction.
This solution is mapped in Poincar\'e patch to a falling monopole
configuration, which is dual to a local quench
triggered by the injection of a condensate.
Choosing boundary conditions
which are dual to a time-independent Hamiltonian, we find
the same functional form of the energy-momentum tensor as the one of a quench 
dual to a falling black hole. 
On the contrary, the details of the spread  of entanglement entropy 
are very different from the falling black hole case,
where the quench induces always a higher entropy
compared to the vacuum, i.e. $\Delta S >0$.
In the propagation of entanglement entropy
for the monopole quench, there is  instead  a competition between
a  negative contribution to $\Delta S$ due to the scalar condensate 
and a positive one carried by the freely propagating quasiparticles generated by the energy injection.

\end{abstract}

\end{titlepage}

\section{Introduction}

The AdS/CFT correspondence \cite{Maldacena:1997re,Gubser:1998bc,Witten:1998qj}  
provides a controlled environment to study the physics of a strongly-coupled conformal field theory (CFT)
 on the boundary by dealing with a weakly-coupled gravitational system in the Anti-de Sitter (AdS) bulk. 
 In particular,   it gives a very useful semiclassical picture of entanglement 
 in strongly-coupled systems \cite{Ryu:2006bv,Hubeny:2007xt,Rangamani:2016dms},
 by relating the entanglement entropy of a subsystem in a CFT
  to the area of an extremal surface   anchored at the boundary
of the  subregion in the AdS dual.  Holographic entanglement entropy is also important to understand black holes: for example,
at finite temperature and in the limit of large subregions,  the entanglement entropy is dominated
 by the thermodynamical contribution,  and the Bekenstein-Hawking entropy \cite{Bekenstein:1973ur,Hawking:1974sw}
is recovered. 
 
 The AdS/CFT correspondence provides several setups in which to investigate the thermalisation of
 out of equilibrium systems. Among these, quenches are conceptually simple theoretical settings, describing the evolution of a system triggered
 by a sudden injection of energy or a change of coupling constants.  
 In the case of global quenches, the perturbation is spatially homogeneous
 and, in the bulk, it corresponds to the formation of a black hole.
The evolution of the entanglement entropy has been studied
in several examples of quench protocols, both on the CFT
  \cite{Calabrese:2005in} and on the gravity side
   \cite{Hubeny:2007xt,AbajoArrastia:2010yt,Albash:2010mv,Balasubramanian:2010ce,Balasubramanian:2011ur,
   Buchel:2012gw,Buchel:2013lla,Auzzi:2013pca,Buchel:2014gta,Liu:2013iza,Liu:2013qca}.
While the simplest features of the spread of entanglement can be described by a model
with free quasiparticles   \cite{Calabrese:2005in},  a more detailed understanding 
needs to take into account  the role of interactions \cite{Casini:2015zua}.
 
 A common feature of global quenches is that the energy is
 injected in the whole space. Once some perturbation
 in the entanglement entropy is detected, there is no clear way
 to discriminate the point where it originated. Local quenches allow for a more transparent
 analysis of the spread of quantum entanglement, because the 
 initial perturbation is localised in a finite region of space.
 Moreover, local quenches can be realised in condensed matter systems
 such as cold atoms \cite{Langen:2013,Meinert:2013}  and they may provide a setup in which 
 entanglement entropy might be experimentally  measured \cite{Klich:2008un,Cardy:2011zz}.
 
 
  In CFT, local quenches can be modelled by joining
 two initially-decoupled field theories \cite{Calabrese:2007mtj,Calabrese:2016xau}
 and then evolving with a time-translation invariant Hamiltonian.
 Another approach in field theory is to consider excited states
 obtained by acting with local operators on the vacuum \cite{Nozaki:2014hna,Caputa:2014vaa,Nozaki:2014uaa,Asplund:2014coa}.
In holography, the latter kind of  local quenches  can be described 
by a free falling particle-like object in AdS \cite{Nozaki:2013wia}.
The topic of local quenches was  studied by many authors
both on the CFT and on the gravity side: for example, mutual information 
was considered in \cite{Asplund:2013zba} and finite temperature aspects 
were investigated in \cite{Caputa:2014eta,Rangamani:2015agy,David:2016pzn}.
Bulk quantum corrections were recently studied in \cite{Agon:2020fqs}.
Local quenches obtained  by splitting an initial CFT into two disconnected pieces 
were considered in \cite{Shimaji:2018czt}.

In principle, the arbitrariness of choosing the falling particle on the  gravity side corresponds to
several choices of local quench protocols on the field theory side.
  In a local quench, it is interesting to understand which aspects of the physics are universal
 and which one depend on the details of the quenching protocol.
 This raises the question of how the choice of the falling particle in AdS
 affects the physics of the local quench of the boundary theory.
 The natural candidates which are free of singularities 
 are black holes (BH) or solitons.   The  black hole case was studied by several authors,
 see  e.g. \cite{Nozaki:2013wia,Jahn:2017xsg,Ageev:2020acl}.
In this paper we will explore the possibility in which the falling particle 
is a soliton in AdS$_4$, focusing on the case of the 't Hooft-Polyakov monopole \cite{tHooft:1974kcl}.

 Monopole solutions in global AdS$_4$ have been considered by several authors,
 starting from \cite{Lugo:1999fm,Lugo:1999ai}.
 AdS monopole wall configurations have been studied in  \cite{Bolognesi:2010nb,Sutcliffe:2011sr,Bolognesi:2012pi,Kumar:2020hif}.
Holographic phase transitions for AdS monopoles have
 been investigated in \cite{Lugo:2010qq,Lugo:2010ch,Giordano:2015vsa,Miyashita:2016aqq}.
In this paper,  we will consider the theoretical setting introduced in  \cite{Esposito:2017qpj}.
In this situation, specialising to a multitrace deformation, 
 the monopole in global AdS is dual to a theory with spontaneous symmetry breaking.
 A previous study of such a model was performed  numerically.
In this work, we find an approximate analytic solution for the winding-one monopole,
which includes the first-order backreaction on the metric.

Applying the change of variables introduced in   \cite{Horowitz:1999gf}, we map
the time-independent global AdS$_4$ solution
 to a falling monopole configuration in the Poincar\'e patch.
On the field-theory side, this is dual to a 
local quench in a perturbed CFT, induced by the insertion 
of a condensate which breaks the global symmetries of the theory.
Outside the quench, the $SU(2)$ global symmetry of the CFT 
remains unbroken.

To investigate the field theory dual of the falling monopole,
we  compute the expectation values of various local operators.
Depending on the choice of boundary conditions for the scalar,
 several interpretations are possible on the CFT side.
For Dirichlet or Neumann boundary conditions, the falling monopole
is dual to a CFT deformed by a time-dependent source, which performs a non-zero external
work on the system.
For  a particular choice of the multitrace deformation, 
given in eq. (\ref{multitrace-speciale}), 
 the monopole is dual to a theory with a time-independent Hamiltonian.
In this case, the expectation value of the energy-momentum
tensor has the same functional form as the one in the background of a  falling black hole  \cite{Nozaki:2013wia}.
In other words, the energy density of the quench is not sensitive to the presence of the condensate.

To further characterise the field theory dual of the falling monopole,
we  perturbatively compute the entanglement entropy for spherical regions.
Let us denote by $\Delta S$ the difference of entropy between the excited state and the vacuum.
We find a rather  different behaviour for $\Delta S$
compared to the case of the falling black hole:
for the monopole quench, $\Delta S$
for a region centered 
 at the origin is always negative,
 while in the BH case $\Delta S$ is positive.
The negative sign of $\Delta S$  for the monopole quench
 is consistent with the expectation that the 
formation of  bound states, which are responsible for the condensate at the core of the quench,
corresponds to a decrease of the number of degrees of freedom  \cite{Albash:2012pd}.

The paper is organised as follows. In section \ref{sect:monopole-global-AdS}
we consider a static monopole solution in global AdS and we  find an analytical
solution in the regime of small backreaction. 
In section \ref{sect:falling-monopole} we apply the change of variables 
introduced in \cite{Horowitz:1999gf} to the global AdS monopole.
This trick transforms the global AdS static solution
to a  falling monopole in the Poincar\'e patch, which
provides the holographic dual of the local quench. In section \ref{sect:boundary-interpretation}
 we  compute the expectation value of some local operators, including the energy-momentum tensor.
In section \ref{sect:entanglement-entropy} we study the entanglement entropy for various subsystem geometries.
We conclude in section \ref{sect:conclusions}. Some technical details
are discussed in  appendices.


\section{A static monopole in global AdS }
\label{sect:monopole-global-AdS}

We consider the same theoretical setting as in \cite{Esposito:2017qpj}
which, in global AdS, is dual to  a  boundary theory with a spontaneously broken $SU(2)$ global symmetry.
The action of the model is:
\beq
S=\int d^4 x \sqrt{-g} \left[ \frac{1}{16 \pi \, G} \le R -2 \Lambda  \ri +  \mathcal{L}_M  \right] \, ,
\eeq
where $ \mathcal{L}_M$ is the matter lagrangian 
\beq
\mathcal{L}_M = 
- \frac14 F_{\mu \nu}^a F^{a \, \mu \nu} - \frac12 D_\mu \phi^a D^\mu \phi^a - \frac{ m_\phi^2}{2}  (\phi^a \phi^a)   \, .
\label{Lmatter}
\eeq
We choose  the cosmological constant  and the scalar mass as follows
\beq
\Lambda=-\frac{3}{L^2} \, , \qquad m^2_\phi=-\frac{2}{L^2} \, ,
\label{Lambda-e-mphi}
\eeq
where $L$ is the AdS radius.
In eq. (\ref{Lmatter}),  $F_{\mu\nu}=F_{\mu\nu}^a \frac{\sigma_a}{2}$ denotes
the non-abelian field strength of the $SU(2)$ gauge field $A_\mu=A_\mu^a \frac{\sigma_a}{2}$, i.e. 
\beq
F_{\mu\nu}^a=\p_\mu  A^a_\nu -  \p_\nu  A^a_\mu+ e \, \epsilon^{abc} A^b_\mu A^c_\nu \, ,
\eeq
with $e$ the Yang-Mills coupling.
The covariant derivative acting on the adjoint scalar is
\beq
D_\mu \phi_a = \p_\mu \phi_a  +  e \, \epsilon_{abc} A_\mu^b \phi^c \, .
\eeq
The equations of motion  are:
\bea
&& D^\mu   F^{a}_{\mu \nu} 
- e \, \epsilon^{abc} \phi^b D_\nu \phi^c = 0 \, ,
\qquad
 g^{\mu \nu} D_\mu D_\nu \phi^a
- m_\phi^2 \phi^a=0 \, ,
\nl
&& R_{\mu \nu}- \frac12 R g_{\mu \nu} + \Lambda g_{\mu \nu} =8 \pi G \,  T_{\mu \nu} \, , 
\eea
where $D_\mu$ denotes the combination
of the gravitational and $SU(2)$ gauge covariant derivatives,
and $T_{\mu \nu}$ is the bulk energy-momentum tensor 
\beq
 T_{\mu \nu} =
D_\mu \phi^a  D_\nu \phi^a +F_{a \mu \a} F_{a \nu}^{\,\,\,\,\,  \a} + g_{\mu \nu} \mathcal{L}_M \, .
\label{bulk_T}
\eeq
We first consider the monopole in a global AdS$_4$ background, with metric
\beq
ds^2= L^2 \le -(1+r^2) d \tau^2 +\frac{dr^2}{1+r^2} +
 r^2 (d \theta^2+\sin^2 \theta d \varphi^2) \ri \, .
\label{global-AdS-1}
\eeq
At large $r$ the field $\phi^a$ has the following expansion
\beq
\phi^a = \alpha^a \frac{1}{r^{\Delta_1}}+ \beta^a \frac{1}{r^{\Delta_2}} + \dots \, , 
\eeq
where $\Delta_{1,2}$ are the dimensions of the sources and vacuum expectation values (VEV)
of the global $SU(2)$  triplet of 
operators $\mathcal{O}^a$ which are dual to the scalar triplet $\phi^a$.
For our choice of mass, see eq. (\ref{Lambda-e-mphi}), the dimensions are
\beq
\Delta_1=1 \, , \qquad \Delta_2=2 
\eeq
and both the $\a^a$ and the $\b^a$ modes are normalisable.
For this reason, we can choose among different possible boundary interpretations
of the source and VEV\footnote{The subscript in the operator $\mathcal{O}^a$ refers to its dimension.}:
\begin{itemize}
\item the Dirichlet quantisation, where ${\a}^a$ corresponds to the source
and ${\b}^a$ to the VEV 
\beq
J_D^a = {\a}^a \, , \qquad   \langle \mathcal{O}_2^a \rangle =  \b^a \, .
\eeq
\item the Neumann quantisation, 
 where $-{\b}^a$ corresponds to the source
and ${\a}^a$ to the VEV 
\beq
J_N^a=-  {\b}^a \, , \qquad     \langle \mathcal{O}_1^a \rangle = {\a}^a \, .
\eeq
\item 
the  multitrace deformation \cite{Witten:2001ua,Berkooz:2002ug,Papadimitriou:2007sj,Faulkner:2010gj,Caldarelli:2016nni},
where  $\langle \mathcal{O}_1^a \rangle = {\a}^a$ and
 the boundary dual is deformed by the action term
\beq
S_{\mathcal{F}}= \int d^3 x \sqrt{-h} \, [ J_{\mathcal{F}}^a \, {\a}^a+\mathcal{F}({\a}^a) ]\, , \qquad
J_{\mathcal{F}}^a= -{{\b}}^a - \frac{ \p \mathcal{F}}{\p {\a}^a}\, ,
\eeq
where $\mathcal{F}$ is an arbitrary function.
Imposing $J_{\mathcal{F}}^a=0$, in order to consider an isolated  system, we find the boundary condition
\beq
{\b}^a=- \frac{\p \mathcal{F}}{\p {\a}^a} \, .
\label{zero-source}
\eeq
\end{itemize}
If we use either Dirichlet or Neumann quantisation,
there is no non-trivial monopole solution with zero boundary scalar sources.
Multitrace deformations, instead, allow  finding a monopole solution with 
a zero boundary source (which satisfies eq. (\ref{zero-source}) for an opportune $\mathcal{F}$),
thus in a situation compatible
 with spontaneous symmetry breaking.

\subsection{Monopole solution in the probe limit}

Let us first consider the zero backreaction limit $G \rightarrow 0$.
The monopole solution can be built by a generalisation of  't Hooft-Polyakov  ansatz in global AdS$_4$
 (see e.g. \cite{Lugo:1999fm,Lugo:1999ai,Esposito:2017qpj}):
\beq
\phi^a=\frac{1}{L} H(r) n^a \, , \qquad 
A^a_l=   F(r) r \,  \epsilon^{a i k} n^k \p_l \, n^i  \, ,
\label{monopole-ansatz}
\eeq
where $x^l=(r,\theta,\varphi)$ and $n^k$ is the unit vector on the sphere $S^2$
\beq
n^k=(\sin \theta \cos \varphi, \sin \theta \sin \varphi, \cos \theta) \, .
\label{direzione-enne}
\eeq
The resulting equations of motion are shown in appendix \ref{Appe-eqs-erre}. 
The regularity of the solution at small $r$ requires that
both $H(r)$ and $F(r)$ approach zero linearly in $r$.
On the other hand, at $r \rightarrow \infty$, the choice of boundary conditions
depends on the physics we want to describe on the boundary.
Such a choice is determined in terms of the coefficients $(\a_H,\b_H,\a_F,\b_F)$ 
specified in the expansion of the scalar and gauge fields nearby the boundary:
\beq
H(r) = \frac{\a_H}{r}+\frac{\b_H}{r^2} + \dots \, ,
\qquad
F(r)  = \frac{\a_F}{r}+\frac{\b_F}{r^2} + \dots \, .
\label{boundarty-condition-F-H}
\eeq
We choose $\a_F=0$ in order to describe a theory
with spontaneous breaking of the $SU(2)$ global symmetry, as in  \cite{Esposito:2017qpj}.
Instead, in order to get a background which is different from empty AdS,
we have to look for solutions where 
$\a_H$ and $\b_H$ are generically non-vanishing.
Once that $\a_H$ is fixed, $\b_H$ is determined by the requirement
that the solution is smooth.
In appendix \ref{flux-appendix} we compute the monopole magnetic flux,
which is independent of the boundary conditions expressed in eq. (\ref{boundarty-condition-F-H}).

An analytical solution to eqs. (\ref{sistema-H-F-senza-backreaction}) can be found at the leading order in
the gauge coupling  $e$:
\bea
H(r) &=& \frac{\alpha_H}{r}  \,  \left[ 1-\frac{\tan^{-1} r}{r}  \right] \, ,
\nl
F(r) &=& \frac{ e \, \alpha _H^2}{16 r^3}  \,  \left[  \pi ^2 r^2-4 \left(r^2+1\right) (\tan ^{-1} r )^2-\left(\pi
   ^2-4\right) r \tan ^{-1} r  \right] \, .
\label{H-F-analitico}
\eea
Such a solution entails the following coefficients
\beq
\label{beta-H-analitico}
 \b_H=-\frac{\pi}{2} \a_H \, , \qquad
 \b_F=e \, \a_H^2 \, \frac{12 \pi -\pi^3}{32} \, .
\eeq
The solution (\ref{H-F-analitico})  is also valid  
for generic $e$, provided that $\a_H$ is sufficiently small.

\begin{figure}
\begin{center}
\includegraphics[scale=0.52]{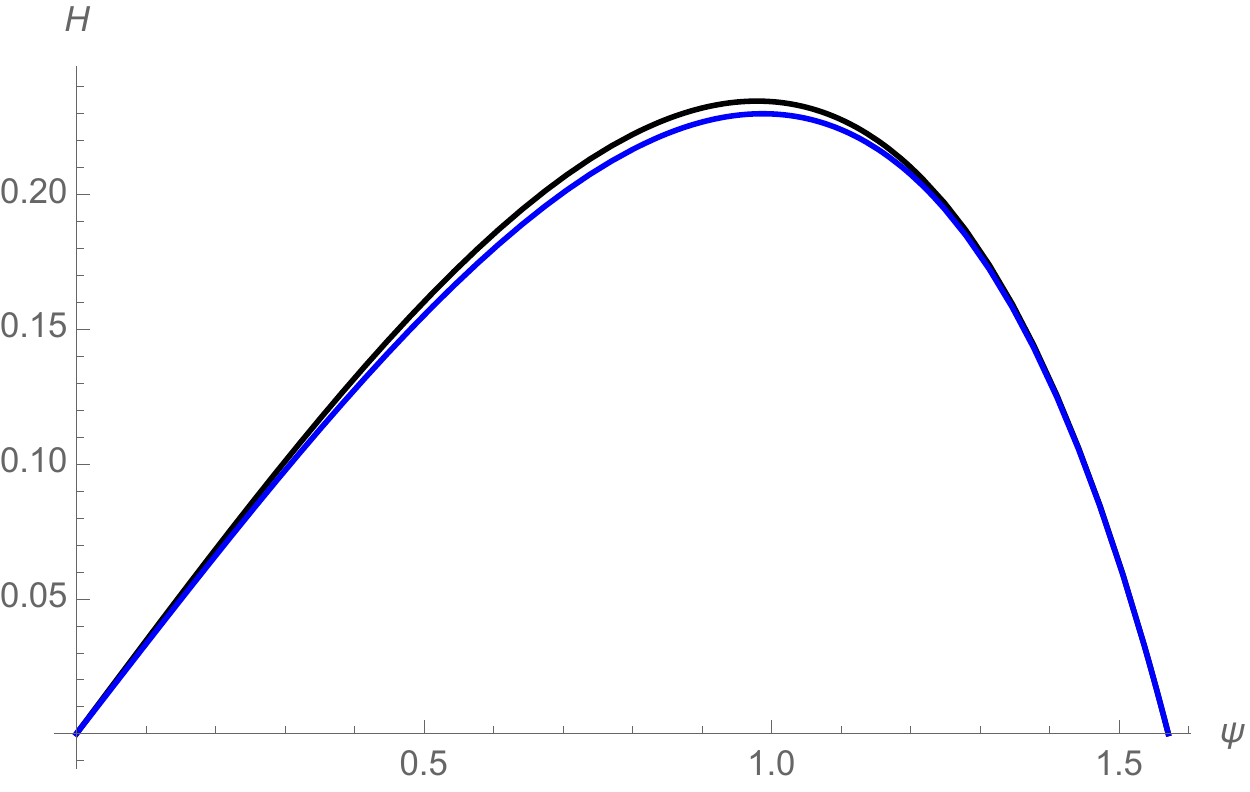} \qquad
\includegraphics[scale=0.52]{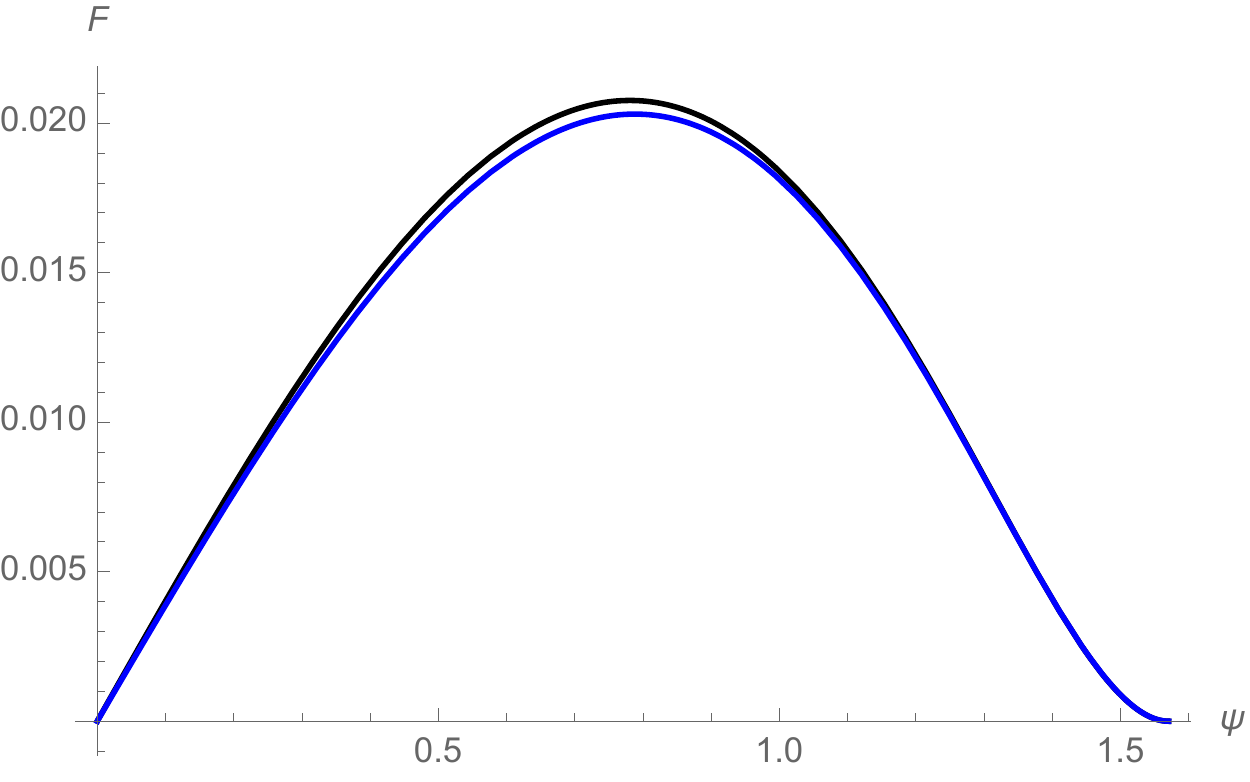}
\caption{
Numerical solutions for $H(\psi)$ and $F(\psi)$ are shown in black
(the values $e=1$,  $\a_H=1$ have been used).
As a comparison, the analytical approximations (\ref{H-F-analitico})
are shown in blue.
}
\label{monopole-profiles-ads}
\end{center}
\end{figure}

\begin{figure}
\begin{center}
\includegraphics[scale=0.7]{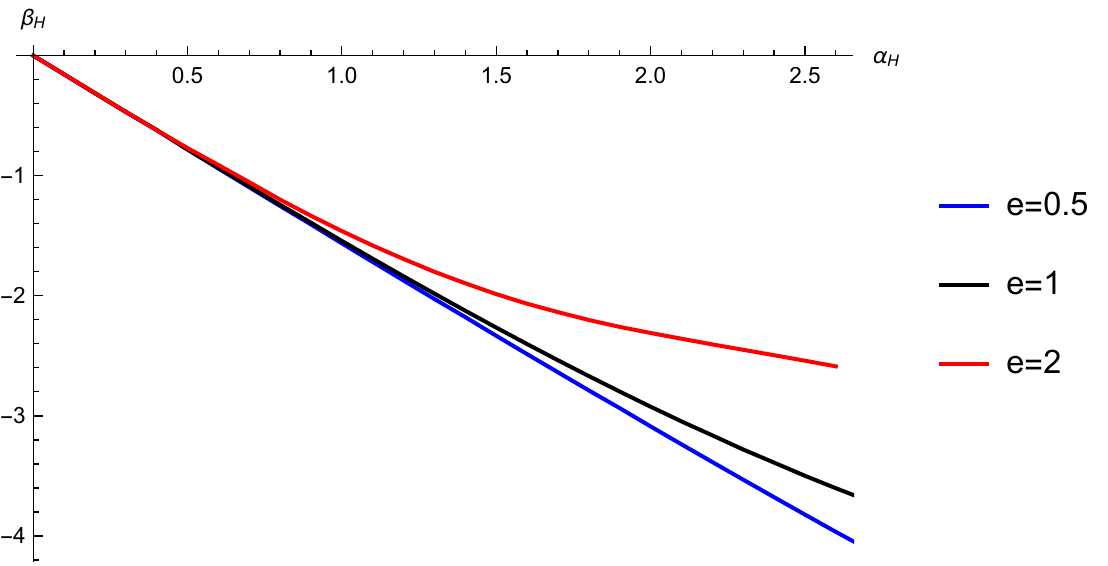} 
\caption{ Plot of the relation between the coefficients $\a_H$ and $\b_H$
for different values of $e$.  The relation in eq. (\ref{beta-H-analitico}) is perturbatively satisfied.
}
\label{alfa-beta}
\end{center}
\end{figure}

At higher order in $e$, eqs. (\ref{sistema-H-F-senza-backreaction})
can be solved numerically. To this purpose, it is convenient
to use the compact variable $\psi$ defined by
\beq
r=\tan \psi \, .
\eeq
The equations of motion in the variable $\psi$ can be found in appendix \ref{Appe-eqs-psi}.
An example of numerical solution is shown in figure \ref{monopole-profiles-ads}.
A plot of $(\a_H,\b_H)$ for various numerical solutions is shown in figure \ref{alfa-beta}.

\subsection{Monopole backreaction}

We now introduce the monopole backreaction, modelled by the metric
\beq
ds^2= L^2 \le -(1+r^2)  h(r) g(r) d\tau^2 +\frac{h(r)}{g(r)} \frac{dr^2}{1+r^2} 
+ r^2 (d \theta^2+\sin^2 \theta d \varphi^2) \ri \, .
\label{monopole-backreaction}
\eeq
In order to recover asymptotically global AdS, we impose the following boundary conditions at large $r$ 
\beq
\lim_{r \rightarrow \infty} h=\lim_{r \rightarrow \infty} g=1 \, .
\eeq
The full set of equations of motion in this background is given in appendix \ref{Appe-eqs-erre}. \\
The asymptotic form of the equations of motion fixes the following large $r$ expansions
\beq
h(r) = 1 + \frac{h_2}{r^2}+\frac{h_3}{r^3} + O(1/r^4) \, ,
\qquad
g(r) = 1 + \frac{g_2}{r^2}+\frac{g_3}{r^3} + O(1/r^4) \, ,
\label{hg-expansion}
\eeq
with 
\beq
g_2=-h_2=\frac{2 \pi G \a_H^2  }{L^2} \, , \qquad h_3= -\frac{16 \pi G}{3 L^2}  \a_H \b_H  \, .
\eeq
The unfixed parameter  $g_3$ can be found by requiring that the solution is smooth.

\begin{figure}
\begin{center}
\includegraphics[scale=0.52]{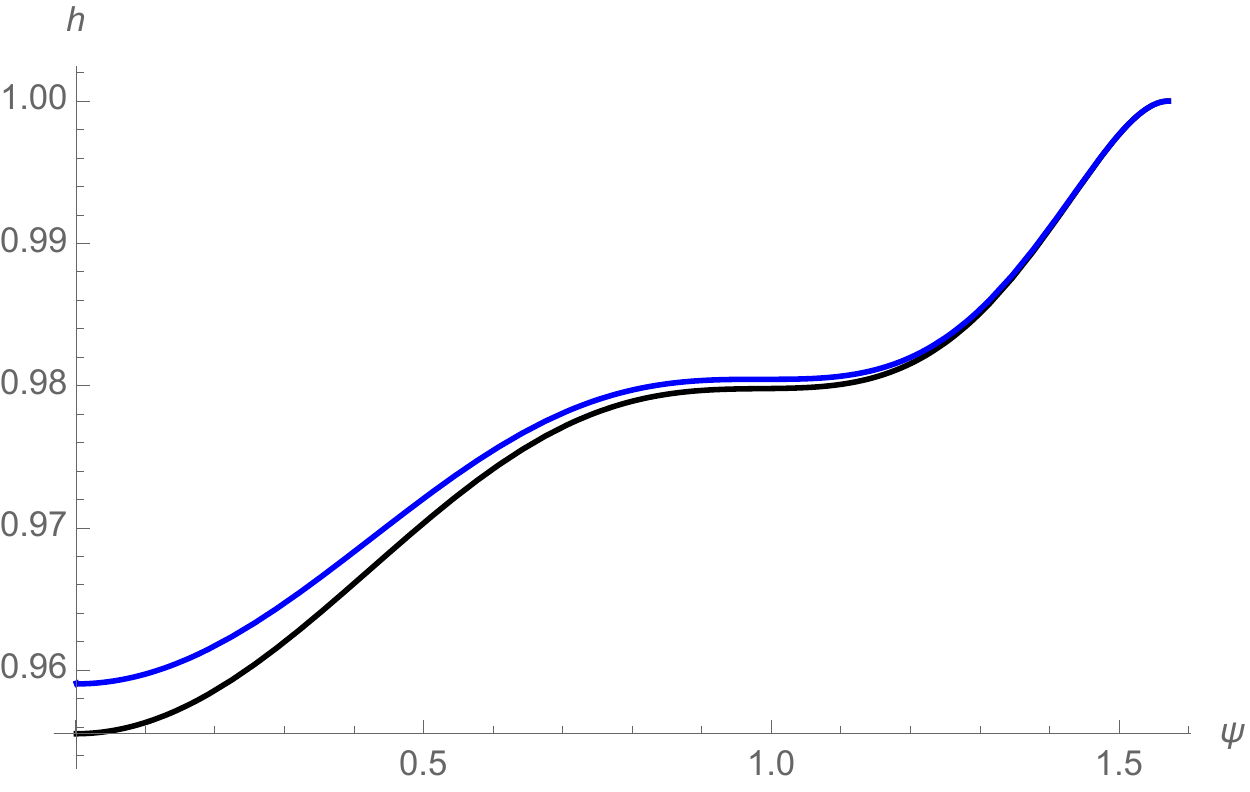} \qquad
\includegraphics[scale=0.52]{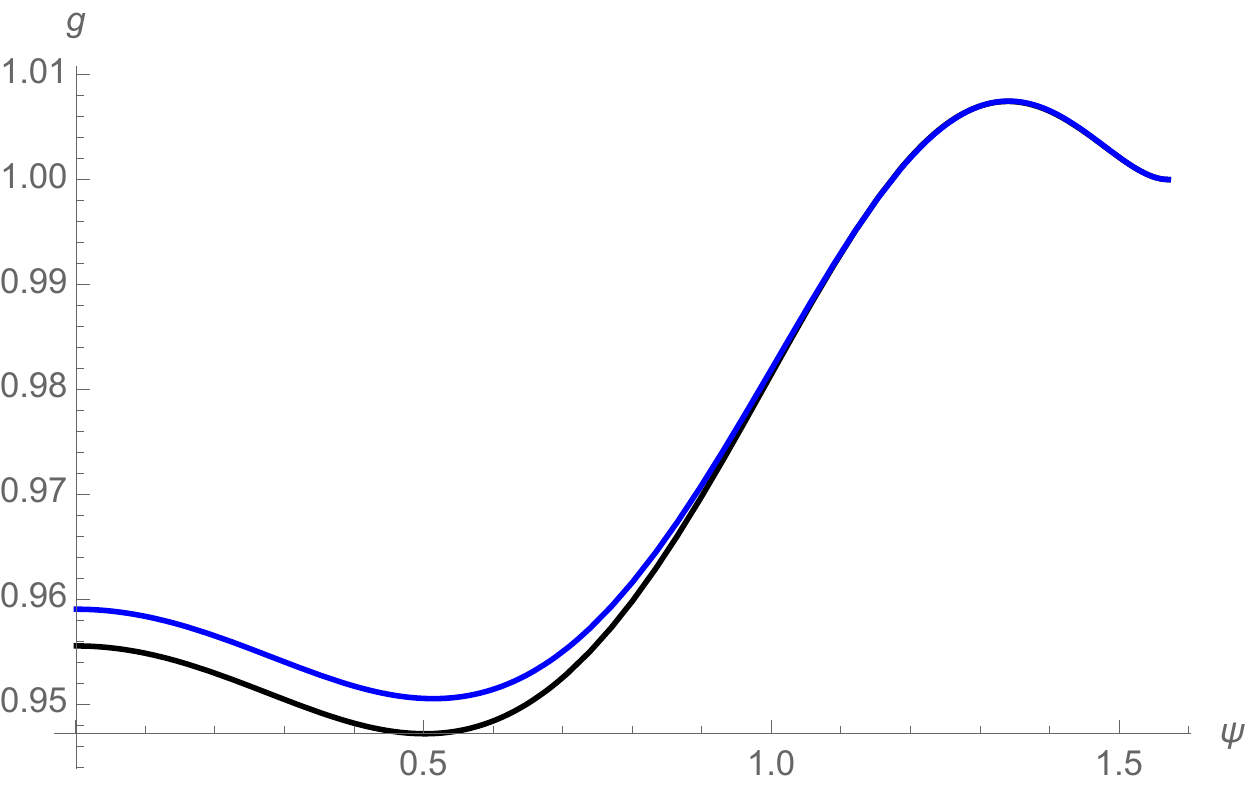}
\caption{
Numerical solutions for the metric functions $h(\psi)$ and $g(\psi)$
are shown in black (the values $e=1$,  $\a_H=1$, $L=1$ and $G=0.1$ have been used).
As a comparison, the analytical approximations (\ref{h-g-analitico})
are shown in blue.
}
\label{monopole-profiles-ads-back}
\end{center}
\end{figure}

In order to treat the problem in an analytical way,
it is useful to introduce the expansion parameter
\beq
\epsilon=\frac{\pi \, G \, \alpha_H^2 }{L^2}  \, ,
\label{epsilon-definition}
\eeq
At the leading order in $e$ and in $\epsilon$, the $H(r)$ and $F(r)$ solutions
are still given by eq. (\ref{H-F-analitico}). The leading order backreaction on the metric
can be solved analytically too, giving:
\beq
h(r)=1 + \epsilon \, h_\ep + O(\epsilon^2) \, , \qquad
g(r)=1 + \epsilon \, g_\ep + O(\epsilon^2) \, , 
\label{definizione-h-g-epsilon}
\eeq
where
\bea
    \label{h-g-analitico}
h_\ep &=&   \pi ^2     -\frac{4}{r^2}  - \frac{2}{r^2+1}
-4 \frac{2 \left(r^2-1\right) r \tan^{-1} r +  \left(r^4+1\right) (\tan^{-1} r)^2 }{r^4}  \, , \nl
\nonumber
g_\ep &=&  \pi ^2 + \frac{1}{r^2} - 
\frac{ 2 r \tan^{-1} r+3  }{ r^2} \left(1-\frac{2}{1+r^2} \right)   \nl
&& 
 -2  ( \tan^{-1} r  ) \frac{  2 \left(r^4-1\right) \tan^{-1} r +r \left(3 r^2+4\right) }{ r^4}
  \, .
\eea
These solutions set
\beq
g_3=-\frac{10 \,  \pi^2 \, G \,  \a_H^2 }{3 \, L^2}  \, .
\label{g3-analitico}
\eeq
The profile functions at higher order in $e$ and $\epsilon$
 are again accessible by numerically solving the equations of motion. 
 As in the probe limit, it is convenient to introduce the variable $\psi =\tan^{-1} r$, 
 getting the equations of motion shown in appendix \ref{Appe-eqs-psi}.
A comparison between the numerical  and the analytical solutions
is shown in figure \ref{monopole-profiles-ads-back}.

\section{A falling monopole in Poincar\'e patch}
\label{sect:falling-monopole}

The gravity dual of a local quench in a CFT can be realised by
considering a falling particle in AdS \cite{Nozaki:2013wia}.
To this purpose, a nice trick was introduced in  \cite{Horowitz:1999gf}.
The idea is to start from a spherically symmetric geometry
in global AdS, and to transform it to a time-dependent
Poincar\'e AdS geometry by performing a change of variables.

The Poincar\'e AdS$_4$ metric with coordinates $(t,z,x,\vphi)$ is
\beq
ds^2=L^2 \le \frac{dz^2 -dt^2 +d x^2 +x^2 d \vphi^2}{z^2} \ri \, .
\label{poincare-AdS}
\eeq
The  metric in eq. (\ref{poincare-AdS}) and the global AdS metric 
in eq. (\ref{global-AdS-1}) can be mapped into each other
 via the coordinate transformations
\bea
\label{A-changing}
 \sqrt{1+ r^2} \cos \tau &=& \frac{A^2+ z^2+ x^2-t^2 }{2  A \,  z} \, , \qquad
 \sqrt{1+ r^2} \sin \tau = \frac{ t}{z} \, ,
\nl
r \, \sin \theta &=& \frac{ x}{z} \, , \qquad  r \, \cos \theta = \frac{   z^2+x^2-t^2 - A^2 }{2 A \, z} \, ,
\eea
leaving the angular coordinate $\vphi$ unchanged. \\
These transformations  can be inverted as follows:
\bea
\label{A-changing-inverse}
r &=& \frac{\sqrt{A^4  + 2 A^2 (t^2+ x^2-z^2) + (z^2 + x^2 -t^2)^2 } }{2 A \,  z} \, ,
\nl
\tau &=& \tan^{-1} \le  \frac{2 A \, t}{ z^2+x^2-t^2 + A^2}  \ri \, ,
\nl 
\theta &=& \tan^{-1} \le \frac{2 A  \, x}{ z^2+x^2-t^2 - A^2 }  \ri \, .
\eea
The change of variables in eq. (\ref{A-changing-inverse}) maps 
a configuration with a static particle in the center of global AdS
to a falling particle in the Poincar\'e patch, that can be used to model a local quench.
We will apply this method to the monopole solution we discussed in section
\ref{sect:monopole-global-AdS}.

The holographic quench is symmetric under time reversal $t \to -t$:
for $t<0$ the monopole is approaching the boundary, while for $t>0$
it moves in the direction of the bulk interior.
Physically, we can think of the initial condition at $t=0$ as 
the initial out-of-equilibrium state, which can be prepared
in the dual conformal field theory by some
appropriate operator insertion.

The position of the monopole center, corresponding to $r=0$ in global AdS,
in the Poincar\'e patch is time-dependent and follows the curve 
\beq
x=0 \, , \qquad z = \sqrt{t^2 +A^2} \, .
\label{traiettoria-monopolo}
\eeq
In the approximation in which the monopole is a pointlike particle,
eq. (\ref{traiettoria-monopolo}) can be interpreted as the trajectory of the monopole.
From the gravity side, the parameter $A$ can be interpreted as the initial position along the $z$-direction
of the free-falling monopole.
From the CFT perspective, the parameter $A$ fixes the size of the local quench.

\subsection{Bulk energy density of the falling monopole}
\label{subsec-energy density and flux}

One may be tempted to imagine the monopole as a pointlike particle
which is falling along the trajectory in eq. (\ref{traiettoria-monopolo}).
To check this intuition, it is natural to consider
the bulk energy-momentum tensor (\ref{bulk_T}). 

\begin{figure}
\begin{center}
\includegraphics[scale=0.45]{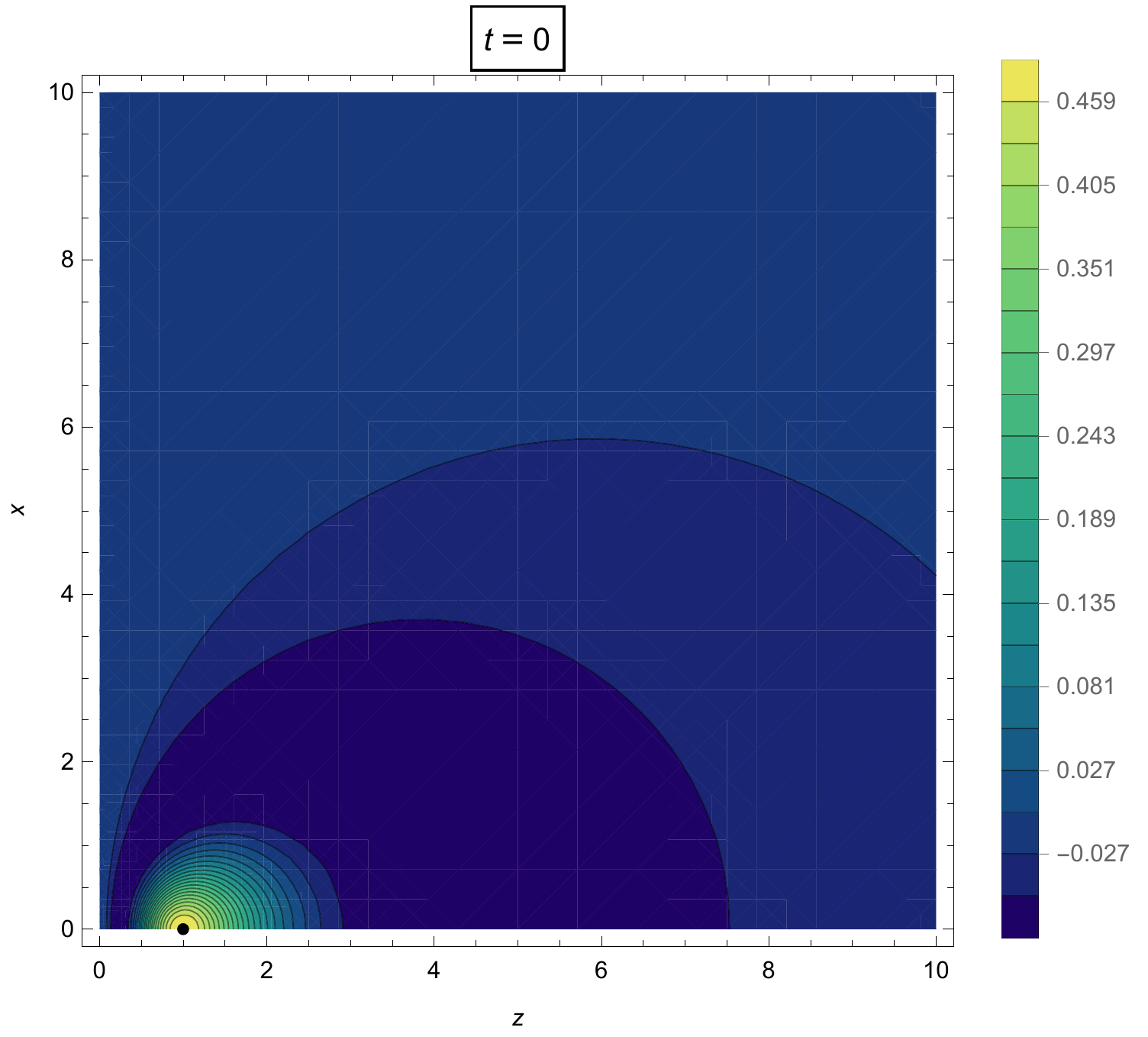} 
\qquad
\includegraphics[scale=0.45]{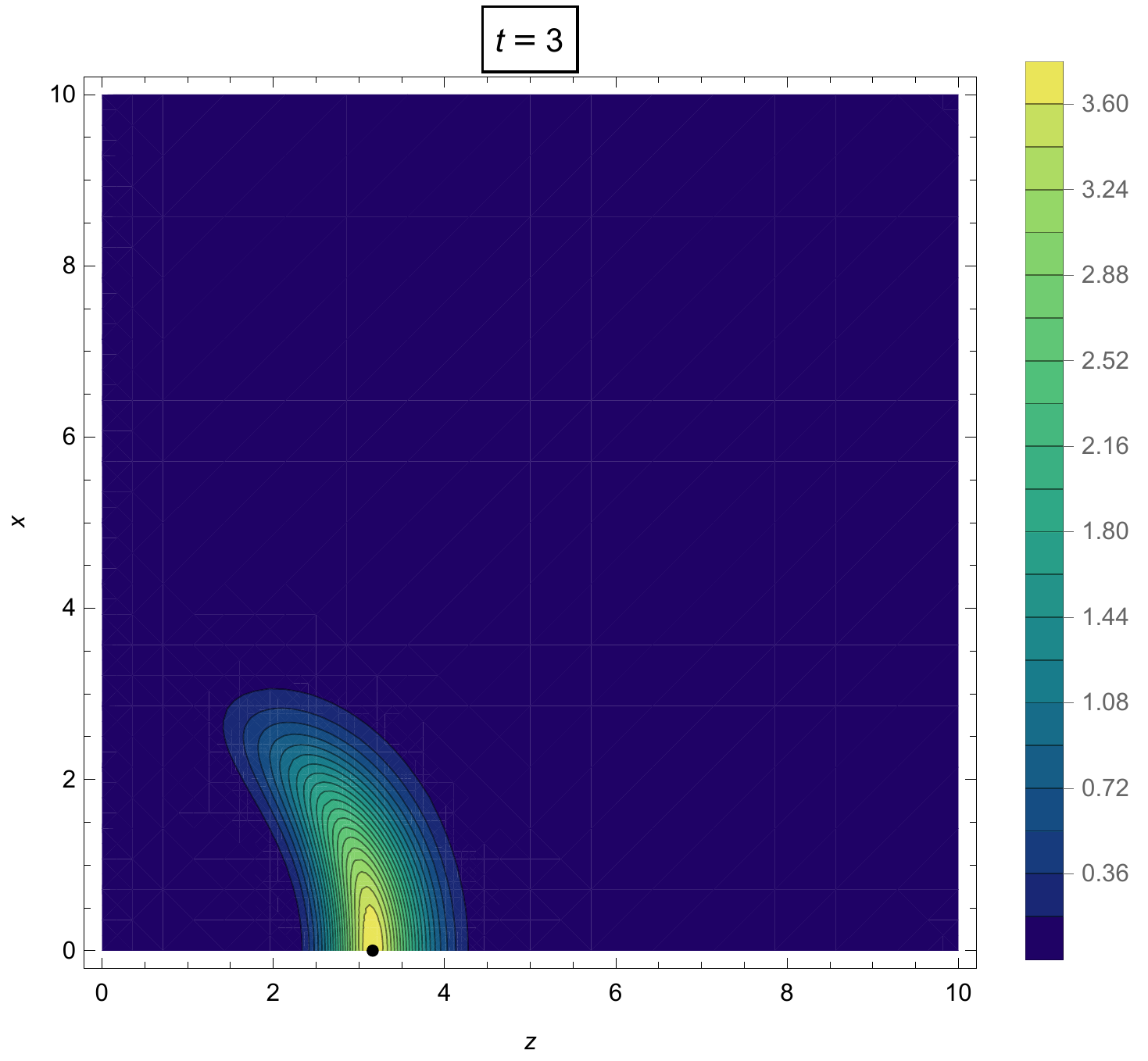}
\qquad
\includegraphics[scale=0.45]{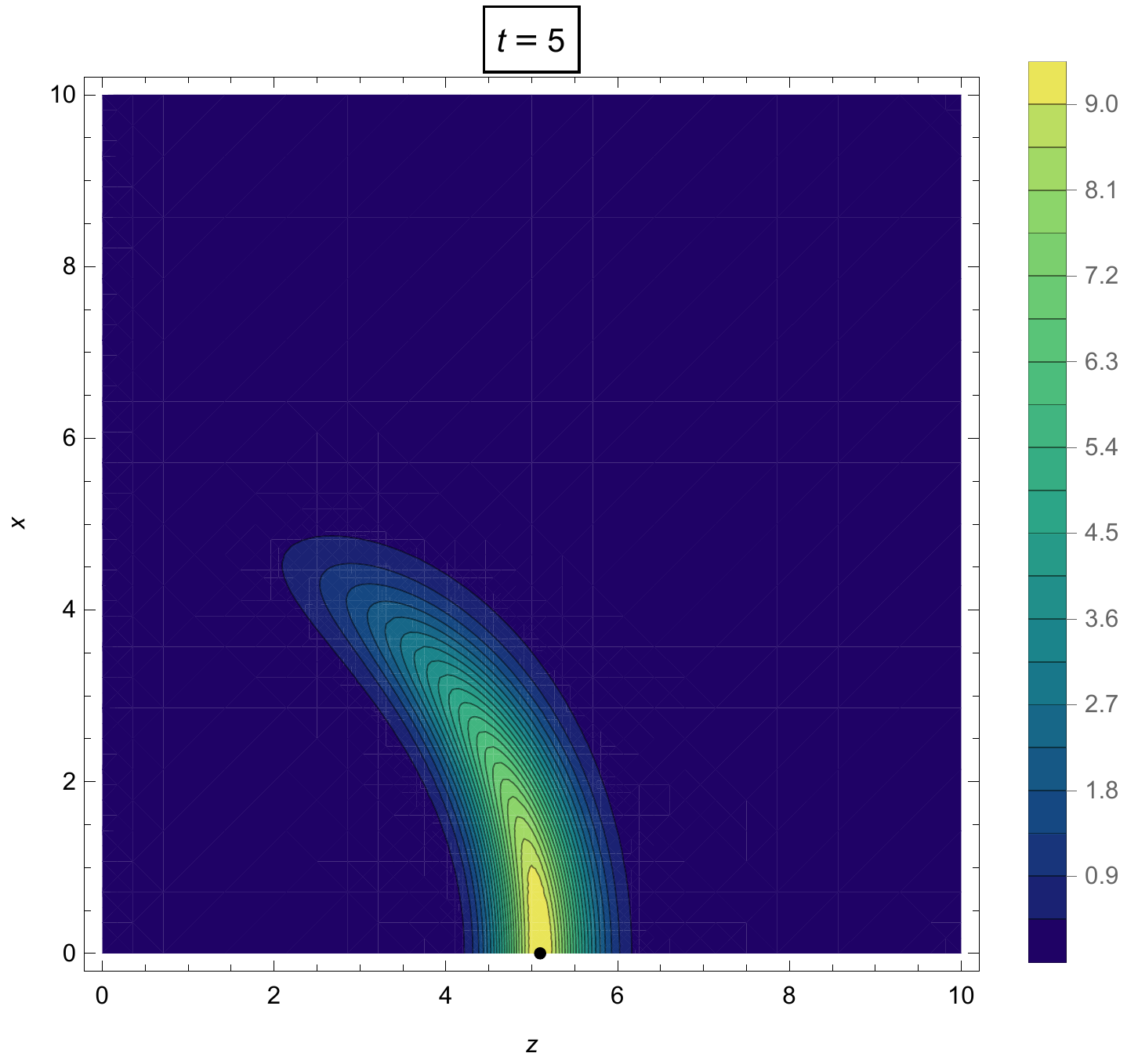}
\qquad
\includegraphics[scale=0.45]{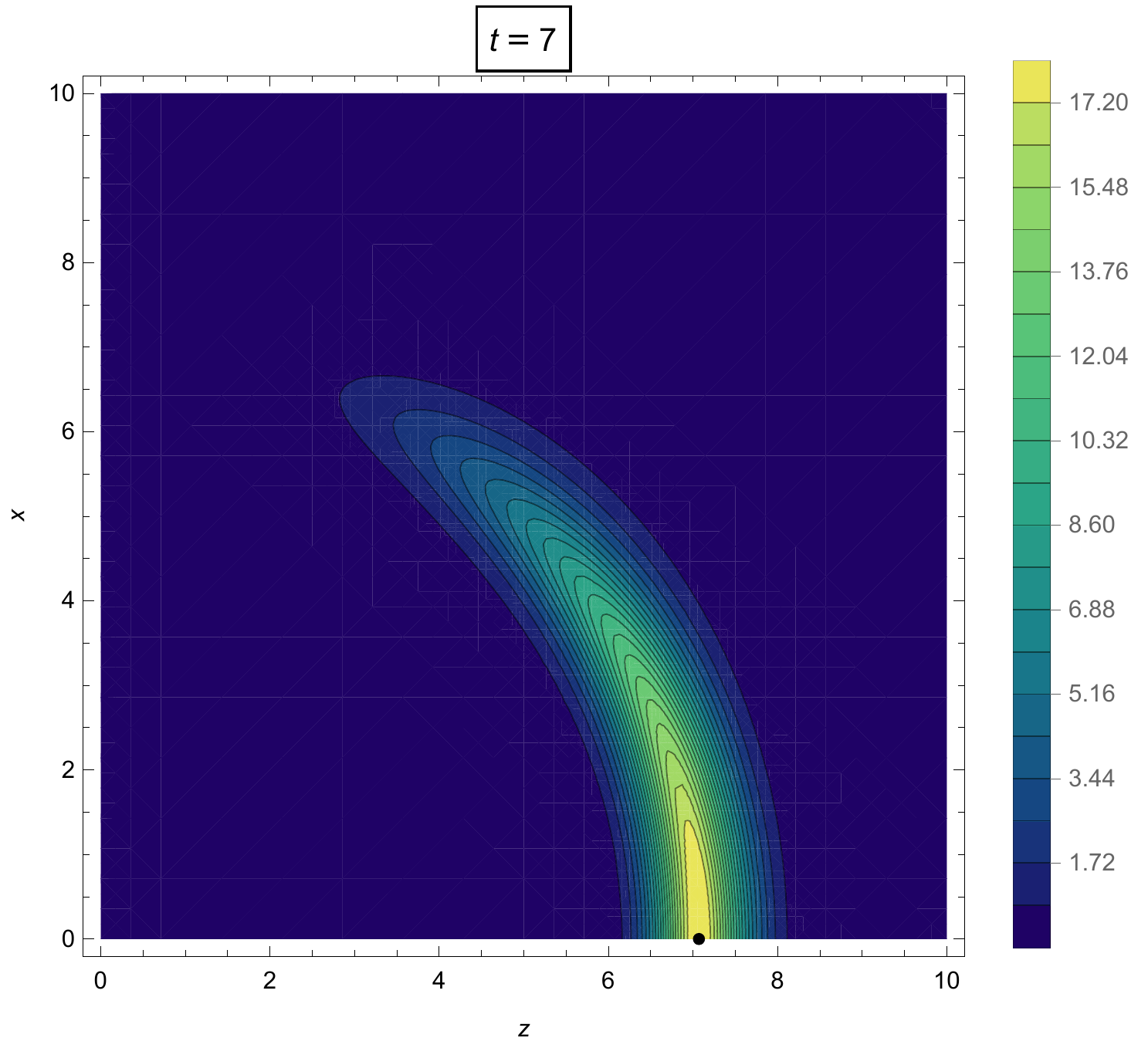}
\caption{Contour lines of constant energy density for fixed time. 
The monopole center is represented by the black spot. 
The numerical values $A=1$ and $L=1$ have been chosen. }
\label{EM00_monopole-plot}
\end{center}
\end{figure}

\begin{figure}
\begin{center}
\includegraphics[scale=0.45]{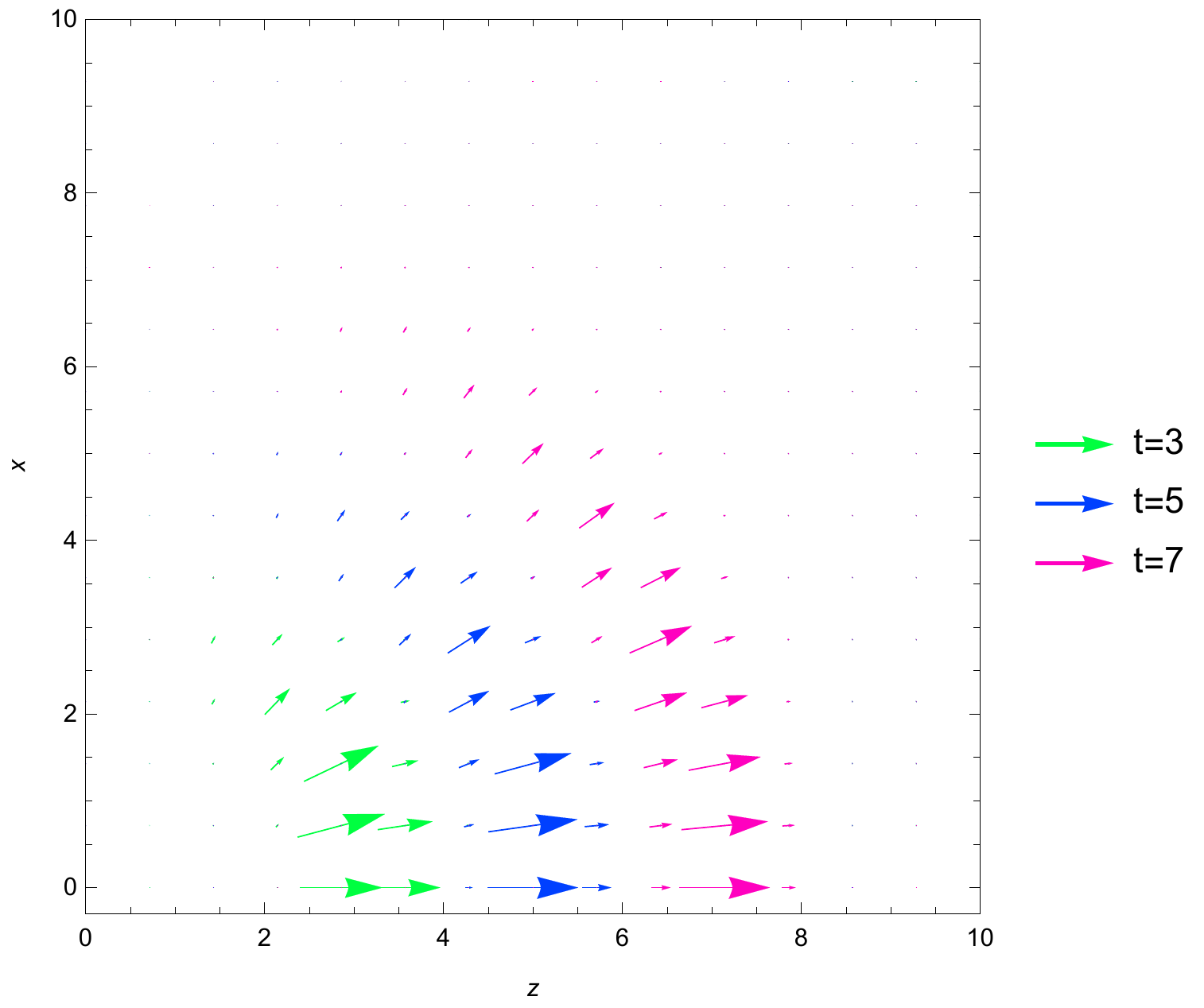}
\caption{Direction of the bulk Poynting vector for fixed time. 
The  numerical values $A=1$ and $L=1$ have been chosen.}
\label{EMpointing_monopole-plot}
\end{center}
\end{figure}

Working in the limit of negligible monopole backreaction,
we perform the coordinate change in eq. (\ref{A-changing-inverse})
\beq
x^{\mu} = \le \tau \, , r \, , \theta \, , \varphi \ri \rightarrow x'^{\mu} = \le t \, , z \, , x \, , \varphi \ri
\eeq
The energy-momentum tensor in Poincar\'e patch is given by
\beq
T_{\alpha \beta}' \le x' \ri = \frac{\partial x^{\mu}}{\partial x'^{\alpha}}
 \frac{\partial x^{\nu}}{\partial x'^{\beta}} \, T_{\mu \nu}\le x \ri \, . 
\eeq
To properly normalise the energy-momentum tensor, we introduce the vierbein $e^{\mu}_m$ such that
\beq
T'_{mn} \le x' \ri = e^\mu_m e^\nu_n \, T'_{\mu \nu} \le x' \ri  \, ,
\qquad
e^\mu_m \, e^\nu_n  \,  g'_{\mu \nu} = \eta_{mn} \, ,
\eeq
where $g'_{\mu \nu}$  and  $\eta_{mn}$  are  the Poincar\'e AdS and the Minkowski
 metric tensors, respectively. In particular, we choose\footnote{In this section, 
 the Minkowski indices $m,n$ take the values $0,1,2,3$,
while the curved spacetime indices are $t,z,x,\vphi$.}
\beq
e^\mu_0 = \le \frac{z}{L} \, , 0 \, , 0 \, , 0 \ri \, , \qquad e^\mu_1 = \le 0 \, , \frac{z}{L} \, , 0 \, , 0 \ri \, , \qquad e^\mu_2 = \le 0 \, , 0 \, , \frac{z}{L} \, , 0  \ri \, .
\eeq
The energy density as measured in such an orthonormal frame is
\beq
\rho = T'^{00} = \frac{z^2}{L^2} \, T'_{tt} \, ,
\eeq
and the components of the Poynting vector $\vec{s}=(s_z, s_x,s_\vphi)$ are
\beq
s_z = T'^{01} = - \frac{z^2}{L^2} \, T'_{tz} \, , \qquad s_x = T'^{02} = - \frac{z^2}{L^2} \, T'_{tx} \, , \qquad s_\vphi= 0 	\, .
\eeq
In figure \ref{EM00_monopole-plot} and \ref{EMpointing_monopole-plot}
 we show the numerical results for the energy density and the energy flux into the bulk at fixed time.
 
  The pictures clearly illustrate how the energy density, 
  initially localised near the AdS boundary, spreads into the bulk. 
The energy distribution resembles that of a pointlike particle only at early times,
while at late times the energy is spread  along a spherical wavefront.  
At $t=0$ all the components of the Poynting vector vanish, 
implying that there is no energy flux at initial time.


\section{Expectation values of local operators}
\label{sect:boundary-interpretation}

To understand the physical interpretation on the boundary CFT,
it is useful to study the expectation values of some important local
operators. In particular, in this section we will focus on expectation values
of scalar operators, global $SU(2)$ currents and energy-momentum tensor.
The details depend on the boundary conditions chosen 
for the scalar field $\phi^a$. In general, the boundary energy-momentum tensor $T_{mn}$
is not conserved because the external sources perform work on the system.
With a particular choice of multitrace deformation, see eq. (\ref{multitrace-speciale}),
the system is isolated and $T_{mn}$ is conserved.

\subsection{Boundary conditions for the scalar}

In the AdS/CFT correspondence, the asymptotic behaviour of the bulk fields
is dual to the source and expectation values of operators in the CFT.
For this reason, we will focus on asymptotics of the scalar field $\phi^a$
nearby the boundary.
In global AdS, the direction $n^a$ of $\phi^a$ in the internal $SU(2)$ space is
given by eq. (\ref{direzione-enne}).
In Poincar\'e patch, by performing the coordinate transformation in eq. (\ref{A-changing-inverse})
we find that, nearby the boundary at $z=0$, 
\beq
n^a=\frac{1}{\omega^{1/2}} \, \left(-2 A x \cos (\varphi ),-2 A x \sin (\varphi ),A^2+t^2-x^2\right) 
+O(z^2)  \, ,
\label{enne-poincare}
\eeq
where, for convenience, we introduce the  quantity $\omega(x,t)$
that appears in many subsequent expressions
\beq
\omega(x,t)=A^4+2 A^2 \left(t^2+x^2\right)+\left(t^2-x^2\right)^2 \, .
\label{omega}
\eeq

\begin{figure}
\begin{center}
\includegraphics[scale=0.52]{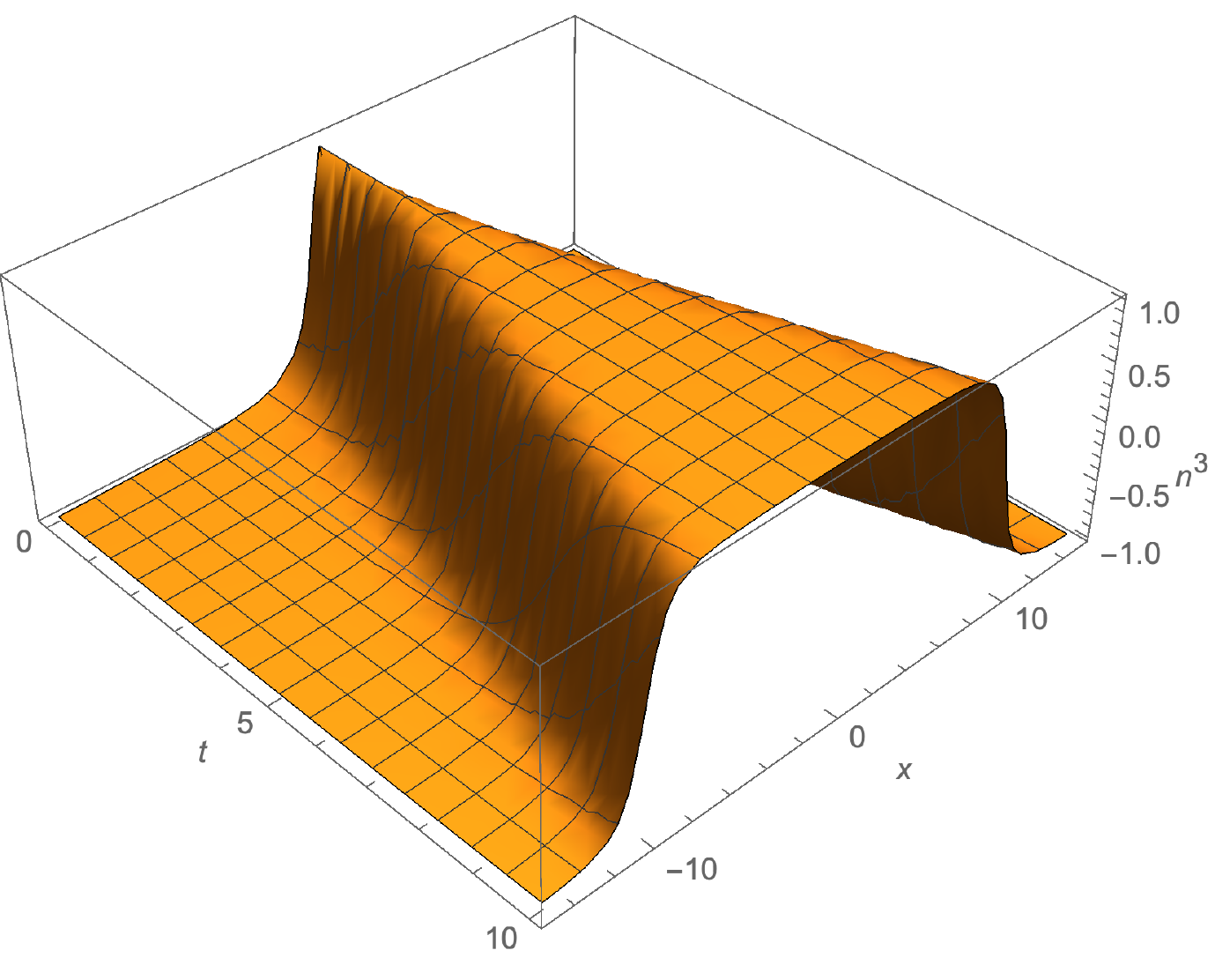} 
\caption{ Value of $n^3$ as a function of $(t,x)$ for $A=1$.
Negative values of the radial cylindrical coordinate $x$ correspond to $\varphi \to - \varphi$.
}
\label{direction-fig}
\end{center}
\end{figure}

The core of the quench can be thought of as localised at
\beq
x=\sqrt{t^2 + A^2} \, ,
\label{cono-luce-largo}
\eeq
which, at large $t$, coincides with good approximation with the lightcone of the origin $x=t$.
For the value in eq. (\ref{cono-luce-largo}),
 the adjoint scalar field points in the direction $n=n^a \sigma^a$ given by
\beq
n=- (\s_1 \cos \vphi + \s_2 \sin \vphi) \, .
\eeq
The scalar points along the $\s_3$ direction inside the lightcone,
and along the $-\s_3$ outside the lightcone, see figure \ref{direction-fig}.
As we will see later, at large $t$, the absolute value of the
scalar field is peaked on $x$ given by eq. (\ref{cono-luce-largo}),
and is almost zero both inside and outside the lightcone. 

 The configuration at $t=0$ resembles a baby skyrmion,
 with a field pointing along $\s_3$ in the core and along $-\s_3$ far away
 (actually, it is not a skyrmion because the VEV tends to zero at infinity).
 As time increases, this configuration expands along the lightcone.
 At large time we end up with two regions of vacuum
 (inside and outside the lightcone) separated by an expanding
 shell of energy.

\begin{figure}
\begin{center}
\includegraphics[scale=0.52]{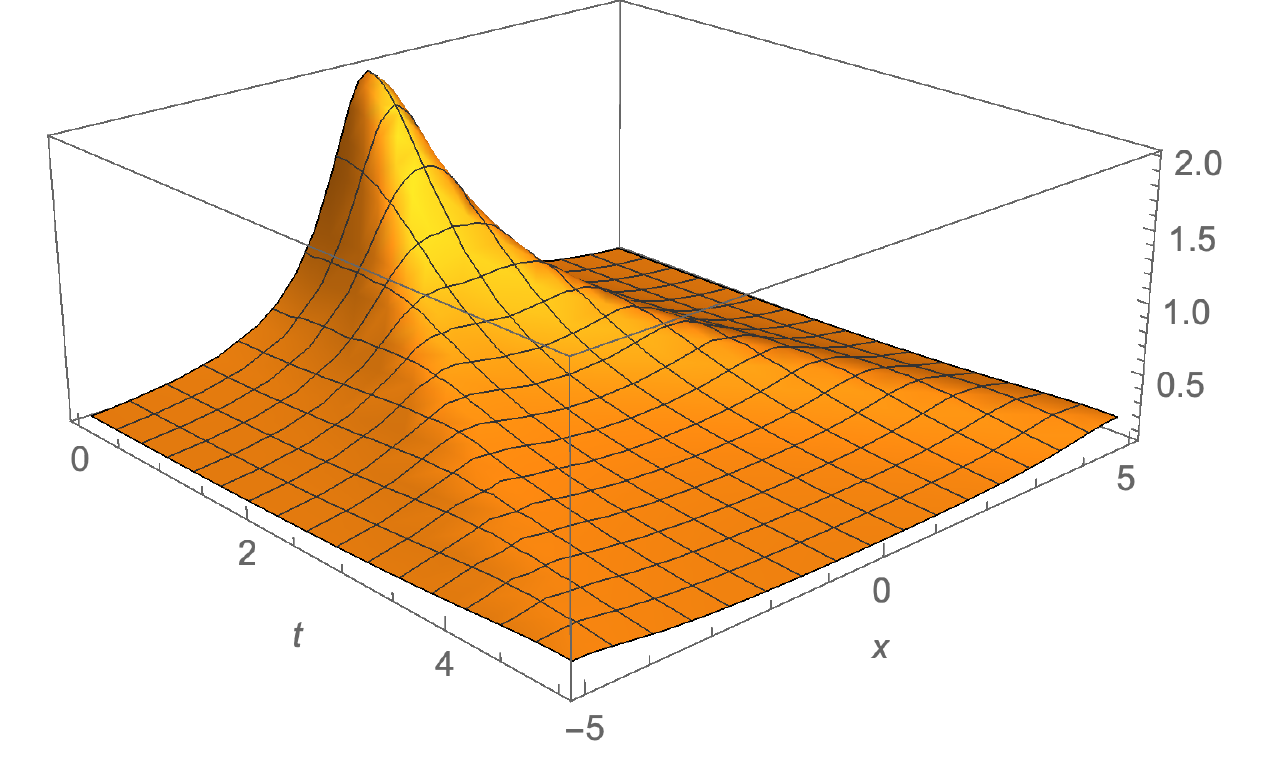}  \qquad
\includegraphics[scale=0.52]{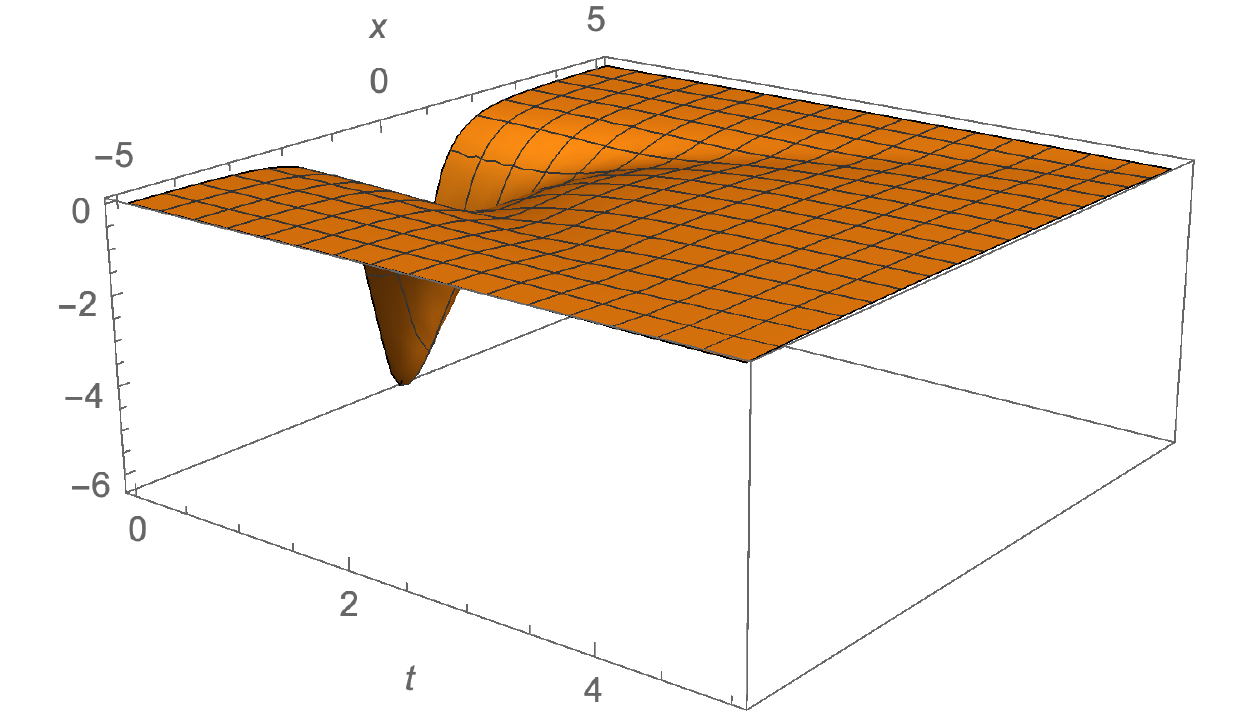}
\caption{ The quantities $\ta_H$ (left) and  $\tb_H$ (right) as a function of $(t,x)$.
Here we set $A=1$, $\a_H=1$ and we use the relations in eq. (\ref{beta-H-analitico}),
 valid for small backreaction. 
}
\label{alpha-beta-tilda}
\end{center}
\end{figure}

In order to extract the sources and the expectation values of the local operator
triggering the quench, it is useful to expand the change of variables in eq. (\ref{A-changing-inverse})
nearby the boundary.
The global AdS radial coordinate reads
\beq
r=\frac{a}{z} +O(z)\, , \qquad
a= \frac{ \omega^{1/2 }}{2 A} \, .
   \label{erre-vs-zeta}
\eeq 
By means of eq. (\ref{erre-vs-zeta}),
we obtain the boundary expansion of $H(r)$
\beq
H=\frac{\a_H}{r} + \frac{\b_H}{r^2}+O(r^{-3})  =
\tilde{\a}_H z + \tilde{\b}_H z^2 + O(z^3)  \, ,
\label{H-poincare}
\eeq
where
\beq
   \tilde{\a}_H =   \frac{\a_H}{a} =    \a_H \,   \frac{2 A}{\omega^{1/2}} \, , 
     \qquad
\tilde{\b}_H =  \frac{\b_H}{a^2}    = \b_H \,  \frac{4 A^2}{ \omega} \, .
   \label{alfa-e-beta-tilde}
\eeq
A plot of $ \tilde{\a}_H $  and $ \tilde{\b}_H $ is shown in figure \ref{alpha-beta-tilda}.
It is interesting to note that 
\beq
\frac{\tilde{\b}_H}{\tilde{\a}_H^2}=\frac{\b_H}{\a_H^2} = \kappa \, ,
\label{alfa-e-beta-tilde-rel}
\eeq
where $\kappa$ is a constant. In the limit of small backreaction, from eq. (\ref{beta-H-analitico})
we find
\beq
\kappa=-\frac{\pi}{ 2 \a_H} \, .
\eeq

Combining eqs. (\ref{enne-poincare}) 
and (\ref{H-poincare}),
 the  expansion of $\phi^a$ nearby the Poincar\'e patch boundary  is
\beq
\phi^a= \frac{H(z)}{L} n^a =\frac{1}{L} \left( {\tilde{\a}}^a \, z + {\tilde{\b}}^a \, z^2 + O(z^3) \right) \, , \qquad
{\tilde{\a}}^a=n^a \tilde{\a}_H \, , \qquad
{\tilde{\b}}^a=n^a \tilde{\b}_H \, .
\eeq
 As for the global AdS case, we can consider several quantisations for the scalar field $\phi^a$:
\begin{itemize}
\item the Dirichlet condition, where $\tilde{\a}_H$ corresponds to the source
and $\tilde{\b}_H$ to the VEV 
\beq
J_D^a  =  \tilde{\a}^a \, , \qquad   \langle \mathcal{O}_2^a  \rangle = \tilde{\b}^a  \, .
\eeq
\item the Neumann condition,  where $-\tilde{\b}_H$ corresponds to the source
and $\tilde{\a}_H$ to the VEV 
\beq
J_N^a= -  \tilde{\b}^a \,  , \qquad     \langle \mathcal{O}_1^a  \rangle = \tilde{\a}^a  \, .
\eeq
\item
the  multitrace deformation, where
 the boundary dual is deformed by the action term
\beq
S_{\mathcal{F}}= \int d^3 x \sqrt{-h}  \, [ J_{\mathcal{F}}^a \, \tilde{\a}^a+\mathcal{F}(\tilde{\a}^a) ]\, , \qquad
J_{\mathcal{F}}^a= -{\tilde{\b}}^a - \frac{ \p \mathcal{F}}{\p \tilde{\a}^a}\, ,
\eeq
and  $\langle \mathcal{O}_1^a \rangle = \tilde{\a}^a$.
\end{itemize}

All these boundary conditions correspond in general
to a monopole in presence of external time-dependent sources.
Among such possible choices of boundary conditions, it is 
interesting to consider the multitrace deformation with 
\beq
\mathcal{F}_\kappa(\tilde{\a}^a) = -\frac{\kappa}{3} \left( \tilde{\a}^a \tilde{\a}^a \right)^{3/2}
=-\frac{\kappa}{3}  \tilde{\a}_H^3 \, .
\label{multitrace-speciale}
\eeq
In this case, the monopole is a solution with a vanishing source, because it satisfies
\beq
\tilde{\b}^a=- \frac{\p \mathcal{F}}{\p \tilde{\a}^a} \, ,
\label{zero-source-poincare}
\eeq
as can be checked from eq. (\ref{alfa-e-beta-tilde-rel}). 

\subsection{The boundary global currents}

Our monopole ansatz in global AdS is given by eq.
(\ref{monopole-ansatz}), with boundary conditions in
eq. (\ref{boundarty-condition-F-H}) and with $\a_F=0$.
As a consequence, we deduce that in Poincar\'e patch the gauge field
$A^a_\mu$ vanishes at the boundary $z=0$.
In other words, if the sources for the global symmetries are set to zero in global AdS,
they also vanish after the change of coordinates leading to the Poincar\'e patch.

\begin{figure}
\begin{center}
\includegraphics[scale=0.52]{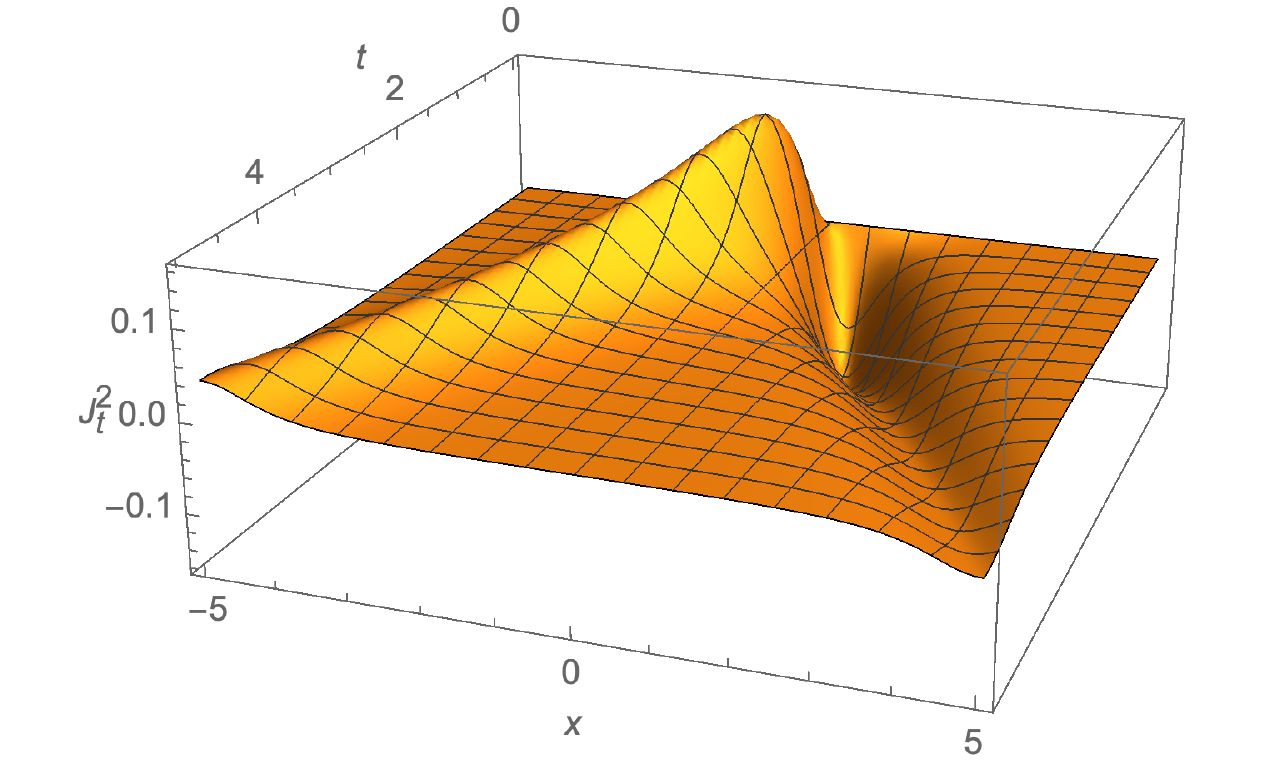} 
\caption{Charge density of the second component of isospin  $J^2_t$ as a function of $(t,x)$
for $\vphi=0$ (positive $x$) and $\vphi= \pi$ (negative $x$).
 We set $A=1$, $\a_H=1$ and $e=1$ and we use the relations in eq. (\ref{beta-H-analitico}),
 valid for small backreaction.
The peak and the pit correspond to positive and negative sign global charges, which are taken apart 
from each other by the quench.
}
\label{density-fig}
\end{center}
\end{figure}

From the order $z$ terms in the boundary expansion of $A^a_\mu$  we can extract the expectation value of the three currents $J^a_l$
\bea
\langle J^1_l \rangle&=& \frac{8 A^2 \beta _F}{\omega^{3/2}}
 \left( t x \sin (\varphi ),-\frac{1}{2} \sin (\varphi )
   \left(A^2+t^2+x^2\right),-\frac{1}{2} x \cos (\varphi )
   \left(A^2+t^2-x^2\right) \right) \, , \nl
\langle J^2_l \rangle &=& \frac{8 A^2 \beta _F}{\omega^{3/2}}
\left( -t x \cos (\varphi ),\frac{1}{2} \cos (\varphi )
   \left(A^2+t^2+x^2\right),-\frac{1}{2} x \sin (\varphi )
   \left(A^2+t^2-x^2\right)\right) \,  ,\nl
\langle J^3_l \rangle &=& \frac{8 A^2 \beta _F}{\omega^{3/2}}
\left( 0,0,-A x^2 \right) \, ,
\eea
where $x^l=(t,x,\varphi)$ are the boundary spacetime coordinates,
and $a=1,2,3$ is the $SU(2)$ global index.
Plots of the charge density $J^2_t$ is shown in figure \ref{density-fig}.

It is interesting to compare the direction in the $SU(2)$ space
of the current expectation value with the direction of the scalar expectation value $n^a$.
We find that the expectation value of the global current is always orthogonal to the direction of the scalar 
expectation value  in the isospin space
\beq
\langle J^a_m \rangle  \, n^a = 0 \, .
\eeq
 So the quench breaks
  all the three global symmetry group generators.
 This is true just on top of the "lightcone" in eq. (\ref{cono-luce-largo}); 
inside and  outside this surface
both the scalar and the current expectation values tend to zero
and  the $SU(2)$ global symmetry of the boundary theory is unbroken.

 \subsection{The boundary energy-momentum tensor}

  For illustrative purposes, we will compare the result with the one obtained for
 a quench modelled by a falling black hole, studied in \cite{Nozaki:2013wia}.
 In this case the metric is
\beq
ds^2= L^2 \le -\le 1+r^2-\frac{M}{r} \ri  d\tau^2 +\frac{dr^2}{1+r^2-\frac{M}{r}} 
+ r^2 (d \theta^2+\sin^2 \theta d \varphi^2) \ri \, ,
\label{metric-BH}
\eeq
where $M$ is a dimensionless parameter proportional to the black hole mass 
\beq
m_{BH}=\frac{1}{2} \frac{M L}{ G} \, .
\eeq
In order to find the quench background, we apply the change of variables in eq. (\ref{A-changing-inverse}).
 Then, to extract the energy-momentum tensor with the method in  \cite{Balasubramanian:1999re},
  it is convenient to pass to Fefferman-Graham (FG) coordinates.  Details of the calculation are
   in appendix \ref{appe-energy-momentum-tensor:BH}, where
 expressions for all the components of the energy-momentum tensor can also be found, see eq. (\ref{T-cono-luce-BH}).
 In particular, the energy density is
 \beq
 T_{tt}^{(BH)}=\frac{A^3 L^2  M }{\pi  \, G } \,\, 
 \frac{ \omega + 6 t^2 x^2 } { \omega^{5/2}} \, .
 \eeq
The energy-momentum tensor is conserved and traceless, and 
the total energy is
\beq
\mathcal{E}^{(BH)}
= \frac{L}{A} m_{BH} \, .
\eeq

 In the falling monopole case a more accurate discussion is needed, since
  the boundary  energy-momentum tensor depends on the details of the boundary conditions of bulk fields.
 We first focus on  Dirichlet boundary conditions  \cite{deHaro:2000vlm}.
  Starting  from the backreacted metric in  eq. (\ref{monopole-backreaction}),
we apply the change of variables in eq. (\ref{A-changing-inverse}).
The intermediate metric expression is a bit cumbersome, so in appendix \ref{appe-energy-momentum-tensor:D} we just specify the coordinates expansion that puts it in the FG form. All the non-vanishing components of the energy-momentum tensor $T_{mn}^{(D)}$ obtained from such a metric are given in eq. (\ref{T-cono-luce-D}). 
The energy density is:
\bea
T_{tt}^{(D)}&=&\frac{A^3 }{3 \pi  G   \, \omega^{5/2}} 
   \left[ 48 \pi  G \alpha _H \beta _H  \, x^2 t^2
+  (8 \pi  G \alpha _H \beta _H-3    g_3 L^2  )  (\omega + 6 t^2 x^2)
\right]  \nl
&=&
 2 \pi  \a_H^2   A^3  \, \frac{   \omega + 2 t^2 x^2    }{       \omega^{5/2}}  \, ,
\eea
where in the second line we have inserted the analytic approximations for small backreaction  in eqs. (\ref{beta-H-analitico})
and (\ref{g3-analitico}). 
In this limit, the total energy is
\beq
\mathcal{E}^{(D)}=   \frac{ \pi^2  \a_H^2   }{A} \le 1- \frac{ 2}{3 }  \frac{t^2}{A^2 +t^2}\ri \, .
\eeq
The energy is a decreasing function of time, meaning that the Dirichlet boundary
conditions absorb energy from the bulk.
The non-conservation of energy motivates the investigation of a different quantisation.

A changing of the quantisation conditions causes a 
shift of $T_{mn}^{(D)}$ by finite parts (see e.g.  \cite{Caldarelli:2016nni}).
Here we specialise to a class of multitrace deformations that do not break the $SU(2)$
global symmetry. 
Assuming that
\beq
\tilde{\a}^a= n^a \tilde{\a}_H \, , \qquad
\tilde{\b}^a= n^a \tilde{\beta}_H \, , 
\eeq
which is true for the monopole, we can
write the source as
\beq
J_{\mathcal{F}}^a = n^a \, J_{\mathcal{F}} \, .
\eeq
As a function $\mathcal{F}$ parameterising the multitrace deformation  we choose  
\beq
\mathcal{F}(\tilde{\a}^a)=\mathcal{F}(\tilde{\a}^a \tilde{\a}^a)=\mathcal{F}(\a_H) \, .
\eeq
The current can be written in terms of $\tilde{\a}_H,\tilde{\b}_H$ as follows
\beq
J_{\mathcal{F}}=-\tilde{\b}_H - \mathcal{F}'(\tilde{\a}_H) \, .
\eeq	
The energy-momentum tensor, (see appendix \ref{appe-energy-momentum-tensor:multitrace} for further details) is 
\beq
T^{(\mathcal{F})}_{ij}= T_{ij}^{(D)}+\eta_{ij} [\mathcal{F}(\ta_H) -\ta_H \tb_H - \mathcal{F}'(\ta_H) \ta_H ] \, .
\eeq

Note that this result also applies to the Neumann conditions, that can be seen as a multitrace deformation with $\mathcal{F}=0$.
If we instead specialise to $\mathcal{F}=\mathcal{F}_\kappa$, see eq.  (\ref{multitrace-speciale}),
 the external source is zero and the energy-momentum tensor is conserved.
 Moreover, an explicit computation reveals that the energy-momentum tensor 
has the same functional form as the one for the falling BH:
\beq
T^{(\kappa)}_{ij}= \frac{m_{M}}{m_{BH}}  \, T_{ij}^{(BH)} 
\,  ,
\qquad  m_M=  \frac{16 \pi G \a_H \b_H-3 L^2 g_3}{6 L G } \, ,
\label{Tkappa-monopole}
\eeq
where $m_M$ is the monopole mass.
Using the analytical values 
 in eqs. (\ref{beta-H-analitico}) and (\ref{g3-analitico}), we find 
 \beq
 m_M=\frac{\pi^2}{3} \frac{\a_H^2}{L}=\frac{\pi}{3}  \frac{L}{G}  \epsilon \, .
 \label{massa-monopolo}
 \eeq
The total energy is
\beq
\mathcal{E}^{(\kappa)}
=\frac{L}{A} m_{M} \, .
\eeq
As apparent from eq. (\ref{Tkappa-monopole}), the energy-momentum tensor 
is not a probe enough precise to distinguish between a falling monopole or a falling
black hole in the bulk. In the next section, we will see that the entanglement entropy
of these two falling objects behaves instead in a radically different way.


\section{Holographic entanglement entropy}
\label{sect:entanglement-entropy}

In this section we will study the effect of
the  leading order backreaction on holographic entanglement entropy.
It is useful to use $\ep$ as defined in eq. (\ref{epsilon-definition}) as 
an expansion parameter. We will find that  the corrections to the entropy
due to the classical backreaction are of order $\a_H^2$, where
\beq
\a_H^2=\epsilon \frac{L^2}{G} \frac{1}{\pi} \, .
\eeq
In the limit of extremely small $\epsilon$, bulk quantum corrections
\cite{Faulkner:2013ana}
 can be of the same order
of magnitude as the ones due to classical backreaction (see for example \cite{Agon:2020fqs}). 
Here we will focus on a regime where the quantum corrections 
are negligible. We first fix an $\ep$ sufficiently small in order to justify
the calculation of the backreaction at the leading order. Then, we  choose
\beq
\frac{L^2}{G} \gg \frac{1}{\epsilon} 
\label{LGepsilon}
\eeq
in such a way that the classical bulk contributions dominate 
over the quantum ones, which are of order $(L^2/G)^0$.
This corresponds to a large monopole mass, i.e. $m_M \gg 1/L$,
see eq. (\ref{massa-monopolo}), which means that the dual local operator
triggering the quench has a large operator dimension, as in \cite{Asplund:2014coa}.
In order to trust our analytical approximation 
for a sufficiently large $\a_H^2$, we need to choose the gauge coupling $e$  sufficiently small,
see the discussion below eq. (\ref{beta-H-analitico}).

In the asymptotically global AdS case, the metric at the leading order in $\ep$ is
\beq
ds^2= L^2 \le -(1+r^2) \left[ 1+ \ep \, (h_\ep +  g_\ep) \right] d\tau^2 + \left[ 1+ \ep \, (h_\ep - g_\ep )  \right]  \frac{dr^2}{1+r^2} 
+ r^2 (d \theta^2+\sin^2 \theta d \varphi^2) \ri \, ,
\label{monopole-backreaction-epsilon}
\eeq
where $h_\ep$ and $g_\ep$ are defined in eq. (\ref{definizione-h-g-epsilon}).
We will be interested in the evolution of entanglement entropy for the quench
in Poincar\'e patch, so we  apply the change of variables in eq. (\ref{A-changing-inverse}),
obtaining a time-dependent background.

The metric tensor can be written as follows
\beq
g_{\mu \nu}=g_{\mu \nu}^{(0)} + \epsilon \, g_{\mu \nu}^{(1)} + O(\ep^2) \, , \qquad \ep=\frac{\pi G \, \a_H^2 }{L^2} \, .
\eeq
Given a codimension-two surface $x^{\mu} \le y^\a \ri$ parameterised with coordinates $y^\a = \le y^1, y^2 \ri $, the induced metric is
\beq
G_{\a \b} = \frac{\p x^\mu}{\p y^\a}  \frac{\p x^\nu}{\p y^\b} g_{\mu \nu} \,  .
\eeq
Such an induced metric
can also be expanded as a power series in $\ep$
\beq
\label{Gk}
G_{\a \b}=G_{\a \b}^{(0)} + \epsilon \, G_{\a \b}^{(1)} + O(\ep^2) \, , \qquad
G_{\a \b}^{(k)} = \frac{\p x^\mu}{\p y^\a}  \frac{\p x^\nu}{\p y^\b} g_{\mu \nu}^{(k)} \, ,  \qquad k=0,1 \, .
\eeq
We can compute the change of area of the Ryu-Takayanagi (RT) surface at the leading order in $\ep$, as in  \cite{Nozaki:2013wia}.
To this purpose, we can expand the determinant of the metric in the area functional.
The first order term of this expansion is
\beq
\Delta \mathcal{A}=\frac{\ep }{2} \int d^2 y \,   \sqrt{G^{(0)}} \, \Tr  \left[ G^{(1)} (G^{(0)})^{-1} \right] \, .
\label{Delta-area}
\eeq
It is important to note that, at first order, it is enough to work with the unperturbed RT surface $x^{\mu} \le y^\a \ri$, which simplifies the computation a lot.
The difference in entropy between the excited state and the vacuum at the leading order
 is proportional to eq. (\ref{Delta-area}) 
\beq
\Delta S = \frac{\Delta \mathcal{A}}{4 G} \, .
\eeq
We will apply this procedure to various examples of subregions.

\subsection{Disk centered at the origin}

We take as a boundary subregion a disk of radius $l$ centered at $x=0$ and 
lying at constant time $t$. The RT surface in unperturbed Poincar\'e patch of AdS$_4$ is  the half sphere
\beq
z=\sqrt{l^2-x^2} \, .
\label{minimal-surface}
\eeq
From eqs. (\ref{Delta-area}) and (\ref{monopole-backreaction-epsilon}) we obtain
\beq
\label{entropy-centrata}
\Delta S(l,t) = \frac{\pi^2 \, \a_H^2}{4} \frac{1}{l} \int_0^l
 \frac{   \left(h_{\epsilon }- g_{\epsilon } \right)  x^3    }
   { \left(l^2-x^2\right)^{3/2}} \, 
     \frac{\omega(l,t) }{    \left(A^2-l^2+t^2\right)^2+4 A^2 x^2 }
    \, dx  \, ,
\eeq
where $\omega$ is defined in eq. (\ref{omega}). The functions $h_{\varepsilon}$ and $g_{\varepsilon}$ depend on the variable $r$, which
on top of the RT surface  reads
\beq
\label{erre-RT-centrata}
r= \frac{ \sqrt{\left(A^2-l^2+t^2\right)^2+4 A^2 x^2 } }{2 A \sqrt{l^2-x^2} } \, .
\eeq

\begin{figure}
\begin{center}
\includegraphics[scale=0.45]{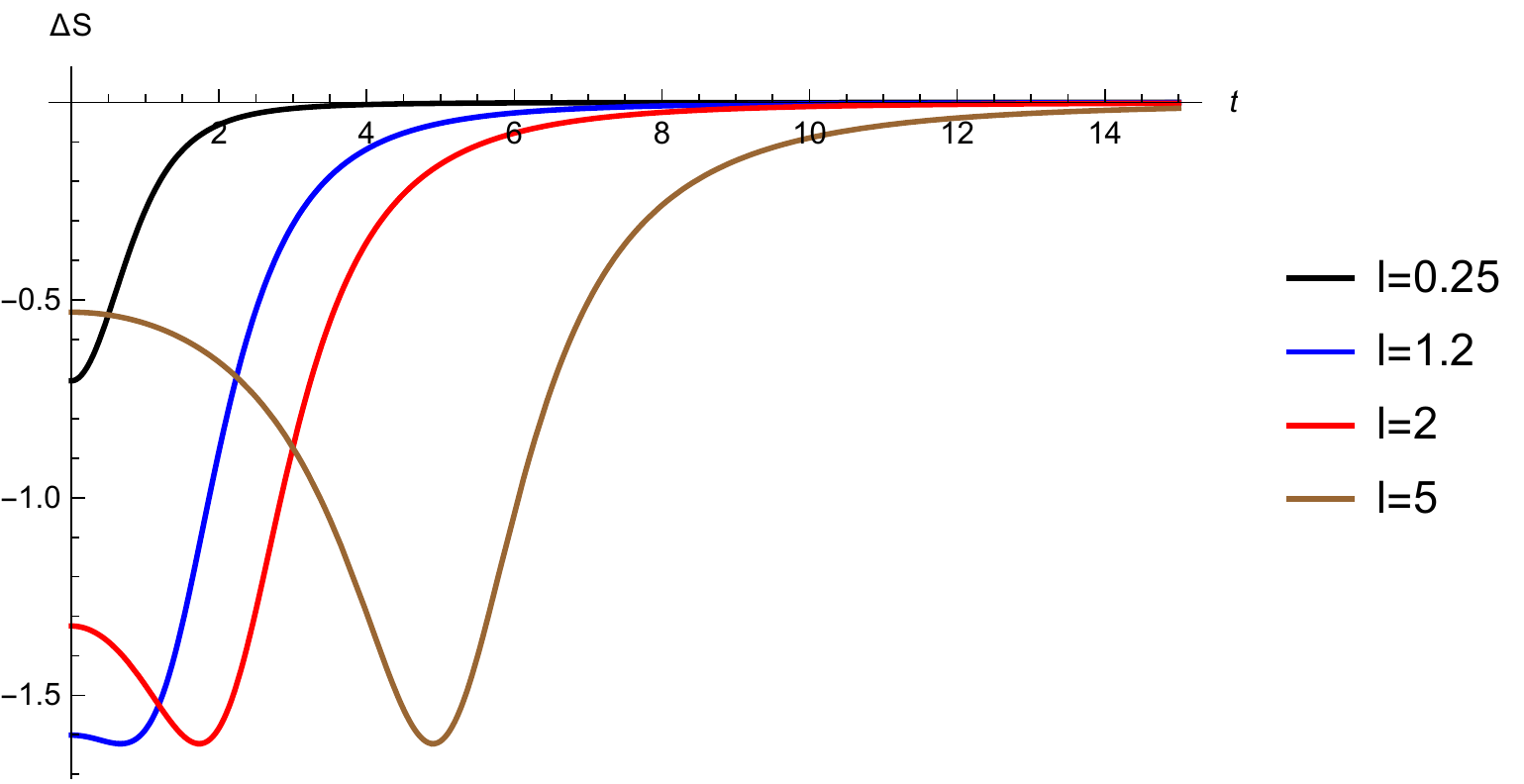}
 \qquad
\includegraphics[scale=0.45]{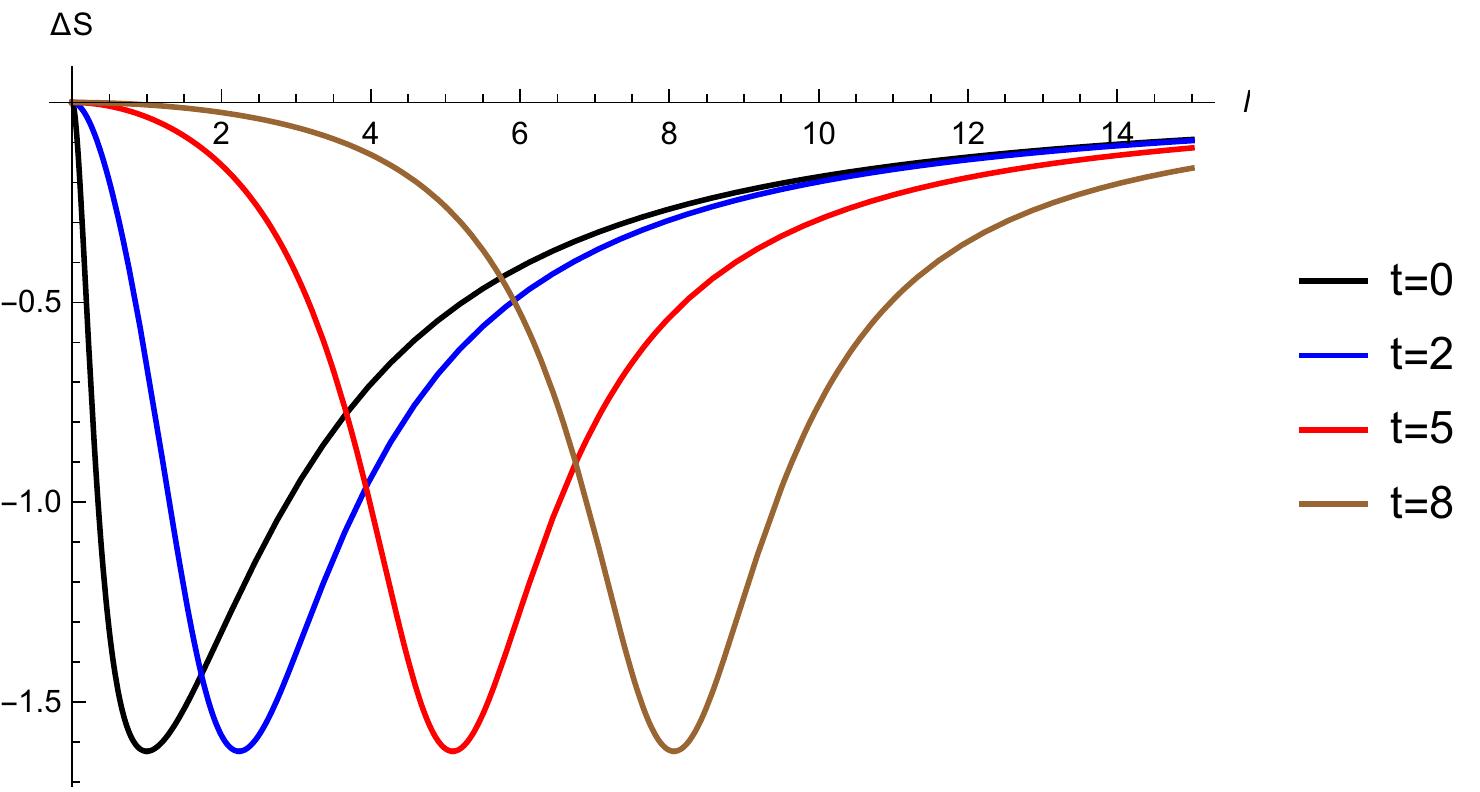} 
\caption{
Left: time dependence of $\Delta S$ for a spherical subregion with fixed radius $l$  centered at the origin of the quench.
 Right: dependence of    $\Delta S$ at fixed $t$ as a function of $l$. The numerical values $\a_H=1$, $A=1$ are used.
 }
\label{entropy-plot}
\end{center}
\end{figure}

For the entropy, the $A$ dependence can be completely reabsorbed
by the following rescaling of the quantities $l$, $x$ and  $t$
\beq
l \to \frac{l}{A} \, , \qquad  x \to \frac{x}{A} \, , \qquad  t \to \frac{t}{A} \, .
\label{rescale-1}
\eeq
For this reason, the numerical analysis has been performed  for $A=1$ without loss of generality.
Numerical results are shown in figure \ref{entropy-plot}.
We find that  $\Delta S$ is always negative,
meaning that the perturbed entanglement entropy is always smaller than the vacuum value.
We can think of the quench  as a region of
spacetime where a condensate (which breaks a global symmetry on the boundary, as in holographic
superconductors \cite{Hartnoll:2008vx}) is localised. A lower entropy
fits with the intuition that some degrees of freedom have condensed \cite{Albash:2012pd}
and so there should be fewer of them compared to the vacuum
(which has zero scalar expectation value).

Analytical results can be found in some regimes. 
Nearby the boundary $r \rightarrow + \infty$, we can use the expansion
\beq
h_{\epsilon } = -\frac{2}{r^2} + \dots \, , \qquad
g_{\epsilon } = \frac{2}{r^2}  + \dots \, .
\label{expansion-h-and-g}
\eeq
Since the minimal $r$ on the RT surface is given by eq. (\ref{erre-RT-centrata})
with $x=0$, this expansion will be valid in the whole integration region in eq. (\ref{entropy-centrata}) in the regime
\beq
{| A^2 + t^2 -l^2 |}\gg 2  l  A\, .
\label{regime-approssimato}
\eeq
Equation  (\ref{entropy-centrata}) can then be evaluated explicitly 
\beq
\Delta S =  -  \pi^2 \, \a_H^2 \,  \left[ \le \frac{A l}{ \sqrt{\omega(l,t)}} +  \frac{ \sqrt{\omega(l,t)}}{4 A l}\ri
\tanh^{-1} \le \frac{2 A l}{\sqrt{\omega(l,t)}} \ri -\frac12
\right] \, .
\eeq
We can specialise the approximation in eq. (\ref{regime-approssimato})
to the following situations:
\begin{itemize}
\item
small $l \ll A$ 
 \beq   
  \Delta S  = - \frac{8}{3} \pi^2 \, \a_H^2 \,   \frac{A^2  \, l^2}{(A^2+t^2)^2} \, .
\eeq
\item large $t \gg A$  and  $t \gg l$ 
 \beq   
  \Delta S  =  - \frac{8}{3} \pi^2 \, \a_H^2 \, \frac{ l^2 \, A^2}{t^4} \, . 
  \eeq
  \item $t=0$ and $l \gg A$
  \beq
  \Delta S=  - \frac{8}{3} \pi^2 \, \a_H^2 \,  \frac{A^2}{l^2} \, .
  \eeq
  \end{itemize}
   
  It is useful to note that, for given $(l,t)$,  the minimal surfaces in eq. (\ref{minimal-surface}) in the Poincar\'e patch
are mapped by eq. (\ref{A-changing-inverse}) to constant $\tau$ surfaces in global AdS.
These surfaces are attached at $r \to \infty$ to a circle
with constant $\theta=\theta_0$, where
\beq
\theta_0(l,t)=\tan^{-1} \le \frac{2 A \, l}{l^2 -t^2-A^2} \ri \, ,
\label{theta-zero}
\eeq
which corresponds to a parallel on the $S^2$ boundary. 
This shows that $\Delta S(l,t)$ is a function just
of the combination in eq. (\ref{theta-zero}). 

The RT surfaces with 
 \beq
l= l_0 = \sqrt{t^2 + A^2}
 \eeq
corresponds to a parallel with $\theta_0 = \pi/2$. 
These surfaces are special, because they lie on the equator of $S^2$.
Due to symmetry,  we conclude that the RT surface at $l=l_0$ 
 has either the maximal or the minimal $\Delta S$.
For the monopole case, we know that $\Delta S$ is  negative 
and close to zero for $t \to 0$ and large $l$.
So we expect that $l=l_0$ is a minimum of $\Delta S$,
as also confirmed by the numerics in figure \ref{entropy-plot}.
  
 For $l=l_0$, $\Delta S$ can be computed exactly:
  \beq
  \Delta S_0 = \Delta S (l_0) = \frac{\pi^2 \, \a_H^2}{4}   \int_0^{\infty}
 \left(h_{\epsilon }- g_{\epsilon } \right)  \frac{r}{\sqrt{1+r^2}}
\, dr = - \Upsilon   \frac{\pi^2 \, \a_H^2}{4} \, ,
\label{Delta-S-zero}
  \eeq
where
\beq
\Upsilon=6 \pi -12 -8 \pi \, \b(2) +14 \, \zeta(3) \approx 0.658 \, .
\eeq
In this expression, $\b(2)\approx 0.916$ is the Catalan constant
and $\zeta$ is the Riemann zeta function.

Summarising, the entropy of the disk with radius $l_0=\sqrt{t^2+A^2}$
remains constant as a function of the time $t$ and equal to the minimum $\Delta S_0$. This can be heuristically justified as follows.
At large $t$, the bound from causality on the speed of entanglement propagation is saturated: 
$\Delta S$, which originated at $t=0$ from a region nearby $x=0$, spreads at the speed of light.
At small $t$, the speed of propagation is smaller, because
at $t \to 0$ also the matter of the quench has zero velocity:
entanglement spreads with matter.

\subsection{Translated disk}

\begin{figure}
\begin{center}
\includegraphics[scale=0.45]{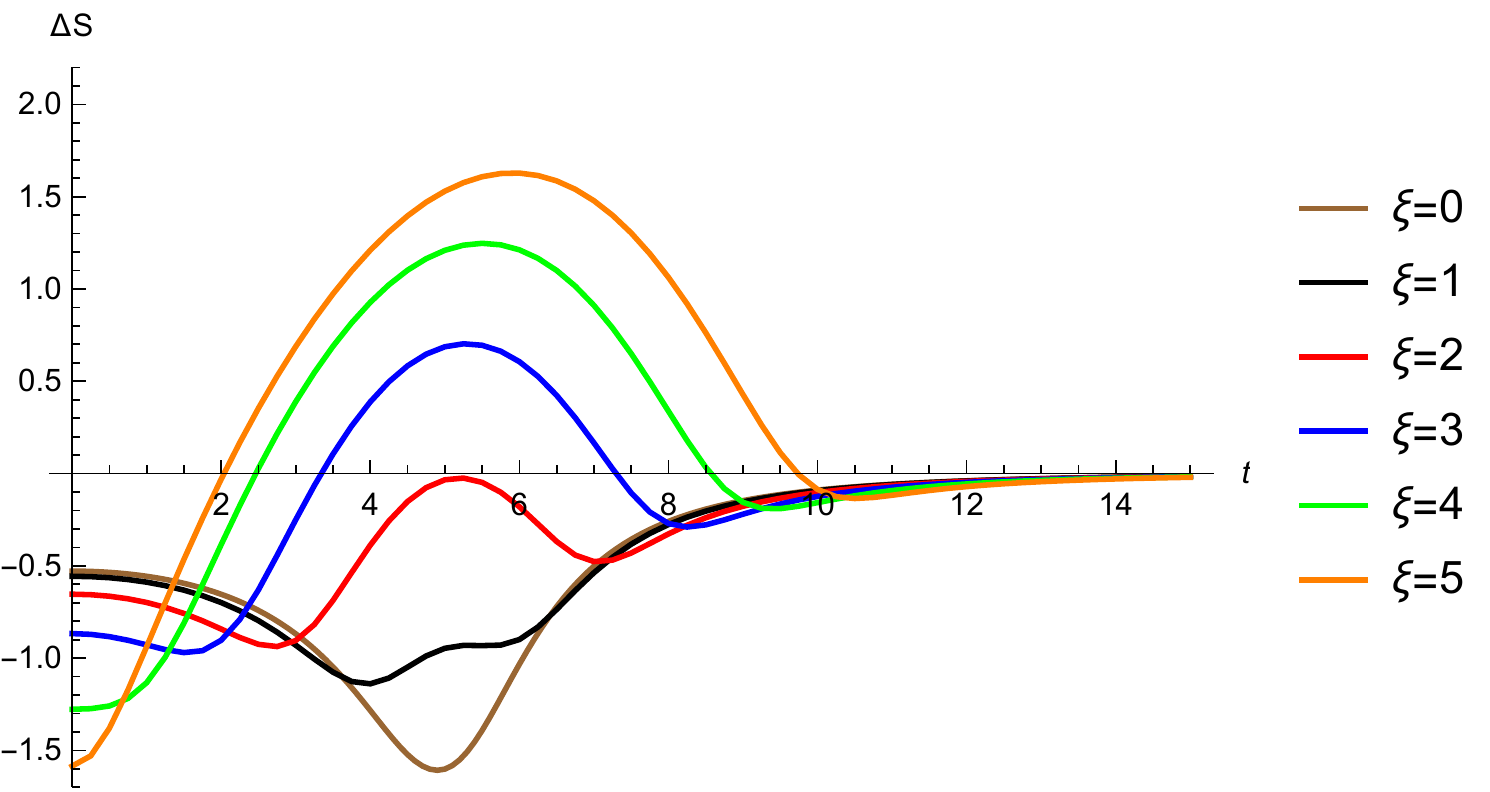} 
\qquad
\includegraphics[scale=0.45]{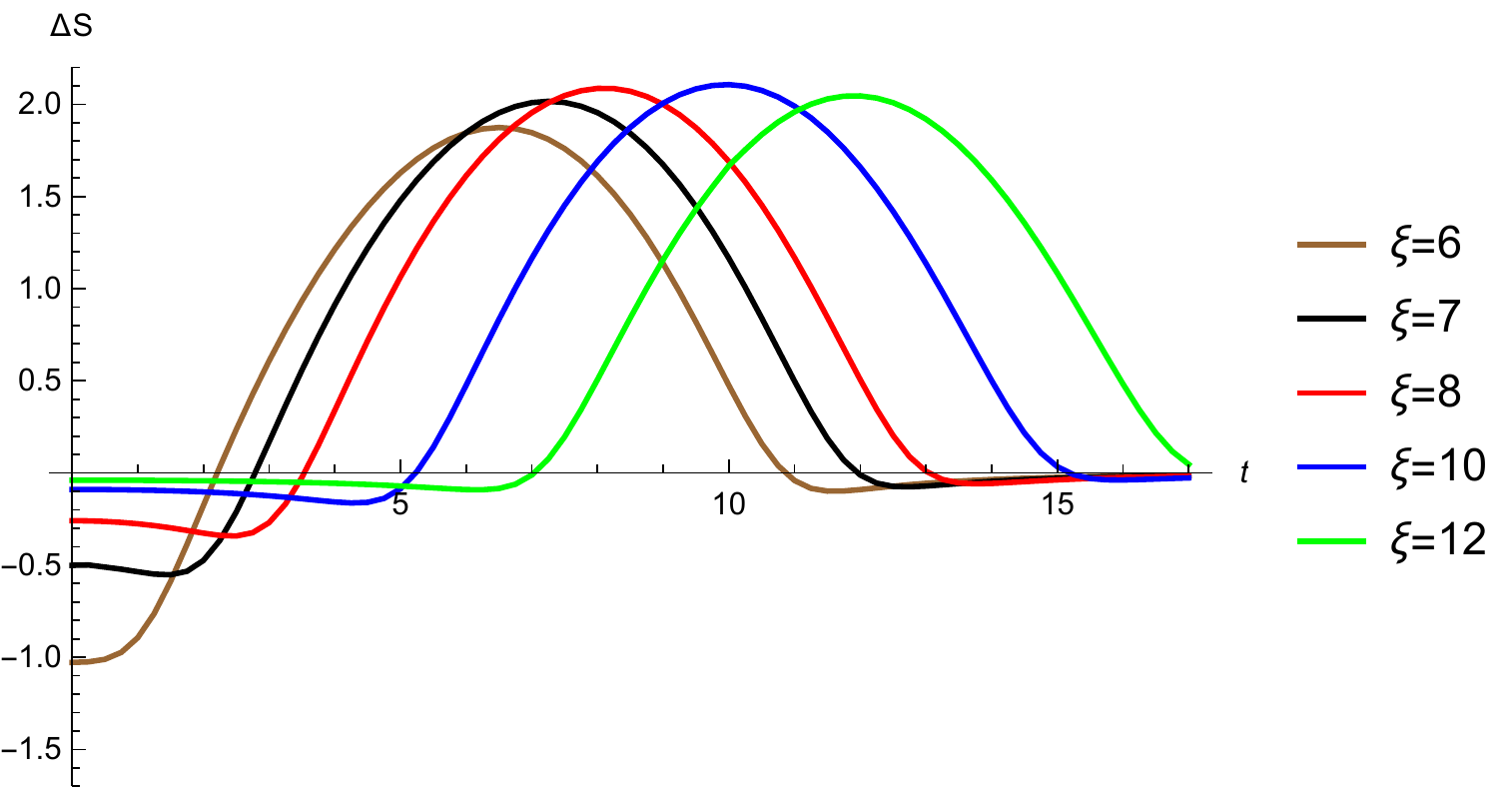}
\qquad
\includegraphics[scale=0.45]{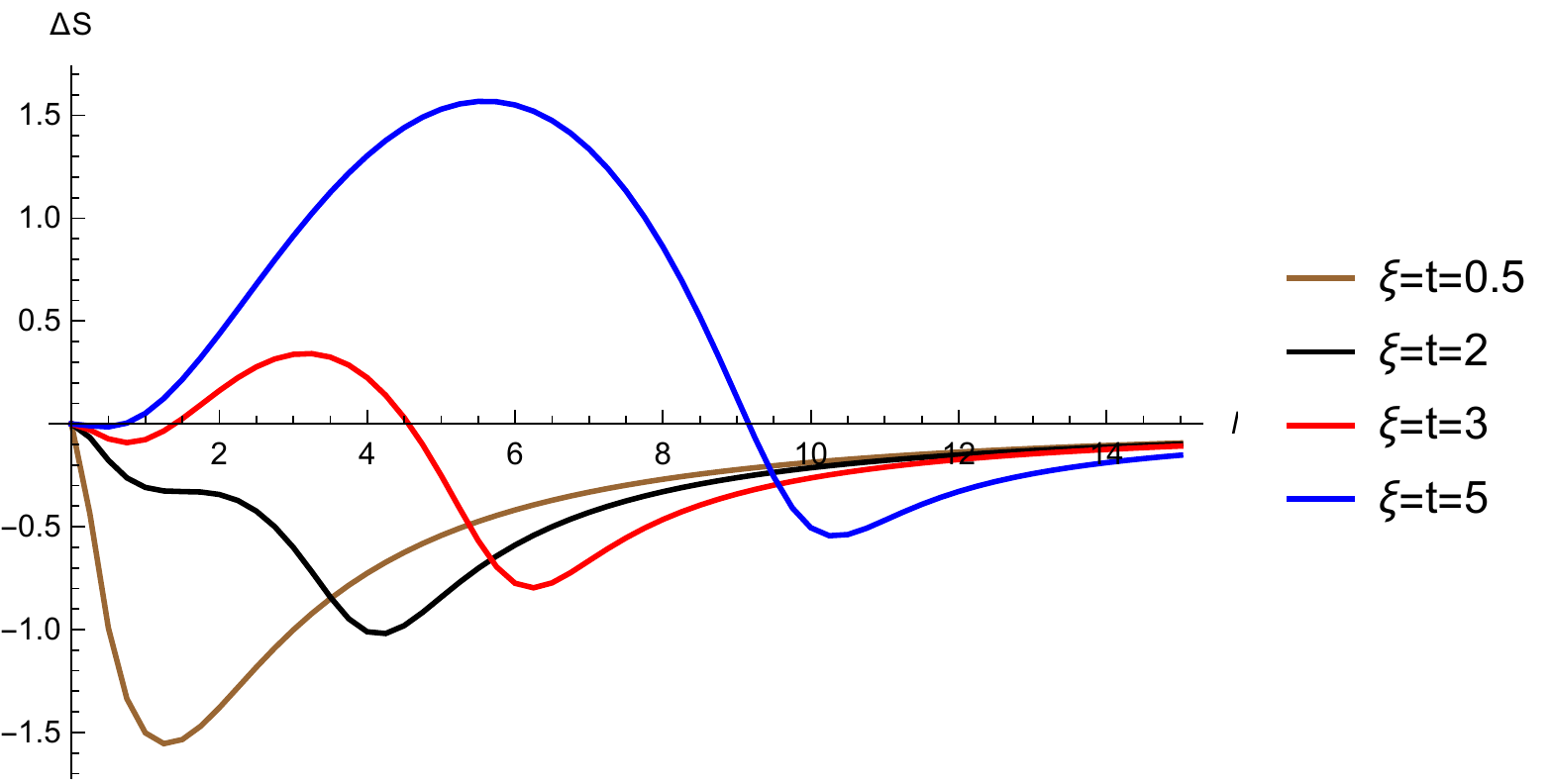}
\caption{
Left and right:  Time dependence of  $\Delta S$ for a disk-shaped subregion of radius $l=5$ centered 
at $\le x_1, x_2 \ri = \le \xi, 0 \ri$ for different values of $\xi$. For large $\xi$, the maximum is reached for $t \approx \xi$.
Bottom: Plot of $\Delta S$ as a function of $l$ for a translated disk-shaped subregion, for various values of $t=\xi$.
Numerical values $\a_H=1$, $A=1$ have been fixed.
}
\label{trans_disk-plot}
\end{center}
\end{figure}

For convenience, we introduce
\beq
 \vec{x} = \le x_1, x_2 \ri=  \le x \cos \vphi, x \sin \vphi \ri\, .
\eeq
We now consider as subregion  a disk of radius $l$ centered at $\le x_1, x_2 \ri = \le \xi, 0 \ri$
  and lying at constant time $t$. The corresponding RT surface in unperturbed Poincar\'e AdS$_4$ is the translated half sphere
\beq
z = \sqrt{l^2 - \le x_1 - \xi \ri^2 - x_2^2} \,  .
\eeq
In appendix \ref{appe:entanglement-entropy-trans-disk} we write the explicit integral for the
first-order correction to the
 holographic entanglement entropy.
It is convenient to rescale spatial and time coordinates as in eq. (\ref{rescale-1}), with $\xi \to \xi/A$ as well.
Numerical results can be obtained for arbitrary radius $l$, see figure \ref{trans_disk-plot}.

In the regime of small $l$, the RT surface stays at large $r$, so we can use the expansion
(\ref{expansion-h-and-g})
and a compact expression can be found 
\beq
  \Delta S(l,t,\xi)
    =  - \frac{8}{3} \pi^2 \, \a_H^2 \,   \frac{A^2  \, l^2}{A^4+2 A^2 \left(\xi ^2+t^2\right)+\left(t^2-\xi ^2\right)^2} \, .
  \label{Delta-S-small-disk}
\eeq
This shows that $\Delta S$ is always negative for disks with small radius $l$.
The entanglement entropy of small disks is then dominated by the negative contribution
due to the scalar condensation.

At large $l \gg \xi$, the subregion is, with good approximation, a
disk centered at $\xi \approx 0$, and so, from the results of the previous
section, we expect a negative $\Delta S$. 
For intermediate $l$, the quantity $\Delta S$ can become positive,
see figure \ref{trans_disk-plot}.
In this regime we can interpret the positive contribution to $\Delta S$ as due to
quasiparticles.

\begin{figure}
\begin{center}
\begin{tikzpicture}
\draw [->] (-4,0)--(4,0) node [at end, right] {$x_1$};
\draw [->] (0,0)--(0,4) node [at end, right] {$t$};
\draw[<->] (-0.02,-0.25)--(0.48,-0.25);
\node at (0.2,-0.6) {$A$};
\draw[red, very thick] (-0.5,0)--(0.5,0);
\draw[blue, very thick] (1.5,2.5)--(3.5,2.5);
\draw[<->] (2.52,2.75)--(3.48,2.75);
\node at (3,3.1) {$l$};
\draw[dashed] (2.5,0)--(2.5,2.5);
\node at (2.5,-0.3) {$\xi$};
\draw[->,red,thick] (0,0)--(2.5,2.5); 
\draw[->,red,thick] (0,0)--(-2.5,2.5); 
\draw[->,red,thick] (0.45,0)--(2.95,2.5); 
\draw[->,red,thick] (0.45,0)--(-1.95,2.5);
\draw[->,red,thick] (-0.45,0)--(-2.95,2.5); 
\draw[->,red,thick] (-0.45,0)--(2.05,2.5); 
\end{tikzpicture}
\end{center}
\caption{In the quasiparticle model, the quench creates EPR pairs of entangled quasiparticles which 
subsequently propagate without interactions. When just one of the quasiparticles belonging to an EPR pair
is inside the blue region, the entanglement entropy of the region increase.}
\label{quasi-particles-figure}
\end{figure}
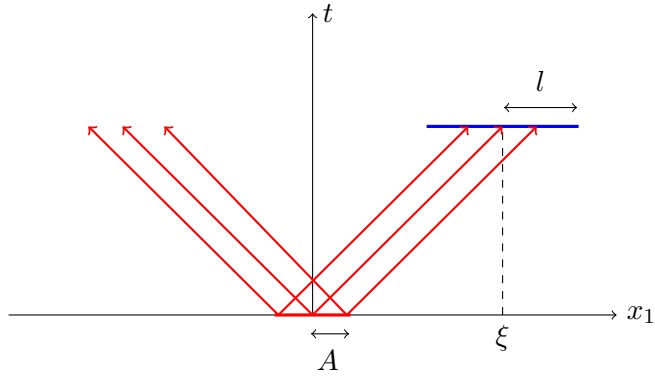

Free quasiparticles provide a simple model of 
entanglement propagation  \cite{Calabrese:2005in}.
In this picture, the quench is assumed to create many copies of 
Einstein-Podolsky-Rosen (EPR)
pairs, which then  propagate without interactions,
see figure \ref{quasi-particles-figure}.
When just one of the entangled particles in an EPR pair
is inside a given region, there is a positive contribution to the entanglement entropy.
This model can reproduce several  aspects of the spread of entanglement
in global and local quenches. Models with interacting quasiparticles
 have also been studied  \cite{Casini:2015zua}.
In all these models, the contribution of the excitations
to the entanglement entropy is always positive.
In the monopole quench, there is also
a negative contribution to the entanglement entropy due to the scalar
condensate. In general, we expect that there is a competition 
between the quasiparticle and the condensate contribution, 
which is responsible for the change of sign of the entanglement
entropy of the translated disk region.

\subsection{Half-plane region}

\begin{figure}
\begin{center}
\includegraphics[scale=0.45]{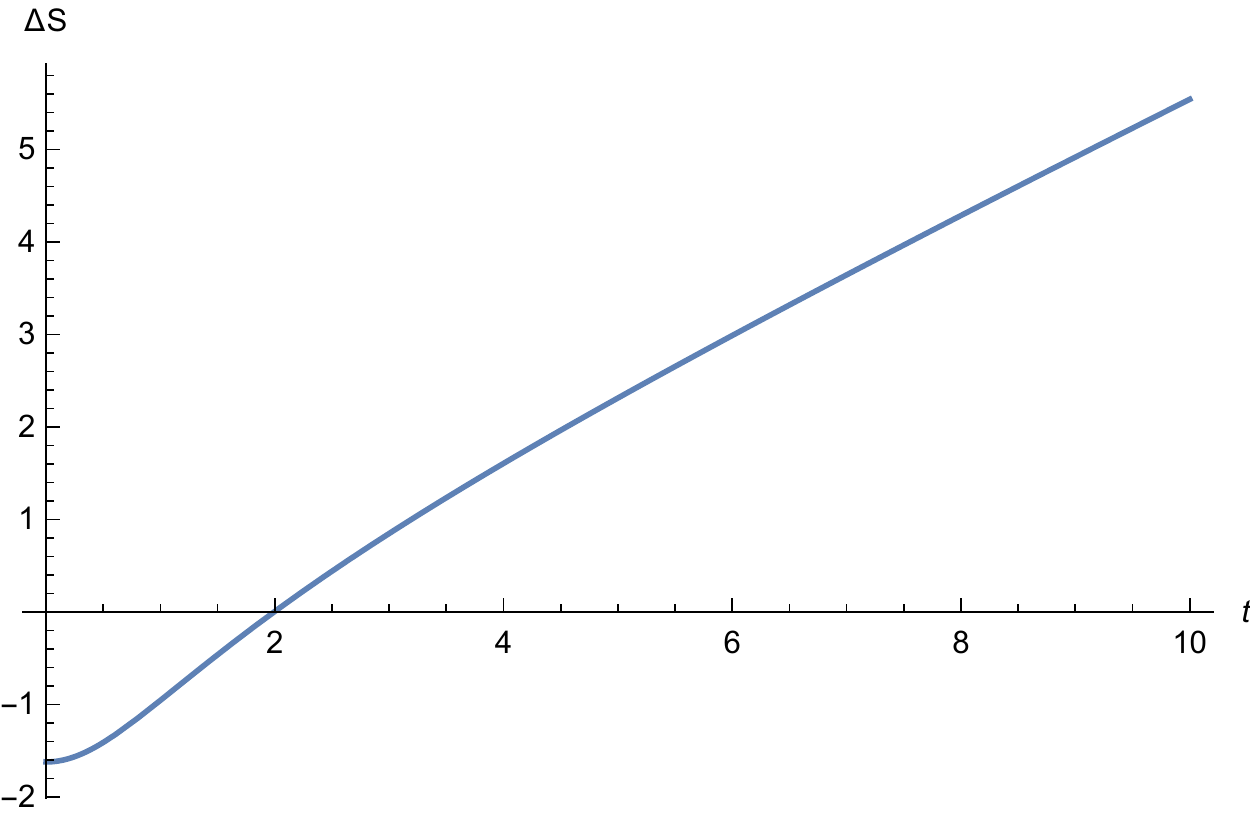}
\caption{
Time dependence of  $\Delta S$ for the half-plane subregion. 
 The numerical values $\a_H=1$, $A=1$ have been chosen.
}
\label{halfplane-plot}
\end{center}
\end{figure}

We take as a boundary subregion the half-plane $x_1 \geq 0$ at constant time $t$. 
The unperturbed RT is the bulk surface at $x_1=0$ and constant time $t$. A convenient choice of parameters is
\beq
y^{\alpha} = \le z, x_2 \ri \, .
\eeq
Details of the calculations are in appendix \ref{appe:entanglement-entropy-half-plane}.
From the closed-form expression, we deduce that the entropy variation $\Delta S$ is a function of $t/A$. 
Numerical result is shown in figure \ref{halfplane-plot}.

For $t=0$, the entropy is given by $\Delta S_0$ in eq. (\ref{Delta-S-zero}).
This is because, due to the change of variables in eq. (\ref{A-changing-inverse}),
the $t=0$ plane with $x_1=0$ (which corresponds to $\vphi=\pm \pi/2$)
is mapped in global AdS to a disk with $\tau=0$ and constant $\vphi=\pm \pi/2$.
Then, an explicit computation easily leads to the same entropy as in  (\ref{Delta-S-zero}).

At large $t$, from the analysis in appendix \ref{appe:entanglement-entropy-half-plane}, we find
that $\Delta S$ scales in a linear way with time, i.e.
\beq
\Delta S = K \, \a_H^2 \,  \frac{ t}{A}  \,  , \qquad K \approx 0.636 \, .
\label{large-time-half-plane}
\eeq
An exact expression for $K$ is given in eq. (\ref{K-expression}).
This is consistent
with numerical results, shown in figure \ref{halfplane-plot}.
We emphasise that this result is valid only in the regime 
where we can trust our perturbative calculation in the parameter $\ep$.
At very large $t$, we expect that the large backreaction effects spoil
the results in eq. (\ref{large-time-half-plane}).
A linear behaviour at large $t$ is also realised for the perturbative entropy
 in the case of a falling black hole in AdS$_4$ \cite{Jahn:2017xsg}.

The numerical plot in figure \ref{halfplane-plot} is consistent
with both the $t=0$ and large $t$ analytical calculations.
At $t=0$, $\Delta S$ is negative, in agreement with the expectation
that the condensate decreases the entanglement entropy.
After an initial transient time,
the quantity $\Delta S$  enters in a linear growth regime
and becomes positive around $t \approx 2 A$.
The asymptotic linear behaviour is similar to the one of the black hole quench
and we expect that it is due to the contribution of quasiparticles.

\subsection{Comparison with the BH quench}

In the case of a falling black hole in AdS$_4$, the perturbative entropy 
for a centered disk is 
\beq
\Delta S^{(BH)}=\frac{\pi M L^2}{4 G \,  A l}\left(\frac{l^{4}-2 l^{2} t^{2}
+\left(A^{2}+t^{2}\right)^{2}}{\sqrt{l^{4}+2 l^{2}\left(A^{2}
-t^{2}\right)+\left(A^{2}+t^{2}\right)^{2}}}-\left|t^{2}+A^{2}-l^{2}\right|\right) \, ,
\eeq
where $M$ is the mass parameter, as defined in eq. (\ref{metric-BH}).
In this case, $\Delta S^{(BH)}$ is always positive
and has a maximum for $l=l_0=\sqrt{t^2+A^2}$.
 The First Law of Entanglement Entropy (FLEE)  \cite{Bhattacharya:2012mi}
is valid in the regime of small $l$:
\beq
\Delta S= \frac{\Delta E}{ T_E} \, , \qquad T_E=\frac{2}{ \pi l} \, ,
\label{FLEE}
\eeq
where $T_E$ is the entanglement temperature and $\Delta E$ is the energy.

FLEE is generically invalidated in the case
of backgrounds with scalars, because the Fefferman-Graham expansion 
of the metric does not start at order $z^d$ \cite{Blanco:2013joa,OBannon:2016exv}, where $d$
is the dimension of the spacetime boundary.
Indeed, as can be checked from eq. (\ref{Delta-S-small-disk}),  FLEE
is not satisfied for the monopole quench background.
The behaviour of $\Delta S$ for small $l$ in eq. (\ref{Delta-S-small-disk}) is rather different from the FLEE regime. In particular:
\begin{itemize}
\item  $\Delta S$ is negative;
\item the quantity $\Delta S$ scales as $l^2$, and not as $l^3$ as predicted by FLEE;
\item there is no  choice of boundary conditions for which
 the energy $\Delta E$ is proportional to $\Delta S$.
\end{itemize}

The  FLEE can be derived from
the notion of relative entropy  \cite{Blanco:2013joa,Rangamani:2016dms},
which is a quantity that measures how far is a density matrix $\rho$ from a reference density matrix $\s$:
\beq
S(\rho || \s) = \Tr (\rho \log \rho) -\Tr (\rho \log \s) \, .
\eeq
As a general property, $S(\rho || \s) $ is positive definite and it
vanishes if and only if $\rho= \s$.
The relative entropy can be written as
\beq
S(\rho || \s) =  \Delta \langle \mathcal{K}_\s \rangle - \Delta S \, ,
\eeq
where $\mathcal{K}_\s$ is the modular Hamiltonian of the density matrix $\s$
\beq
\mathcal{K}_\s = - \log \s \, .
\eeq
From positivity of relative entropy we get the relation
\beq
\Delta \langle \mathcal{K}_\s \rangle  \geq \Delta S \, .
\label{relativeE-ineq}
\eeq
The modular Hamiltonian operator $\mathcal{K}_\s$ for a spherical domain with radius $l$ in the vacuum state of a $d$-dimensional CFT
can be expressed \cite{Casini:2011kv}  in terms of the energy-momentum operator as follows
\beq
\mathcal{K}_\s= 2 \pi \int_{\rm sphere } d^{d-1} \vec{x} \, \,\, \frac{l^2 -x^2}{2 l} T_{tt}(\vec{x}) \,  ,
\eeq
where $x=|\vec{x}|$.
In the limit of small spherical subregion and in $d=3$, we find
\beq
\Delta \langle \mathcal{K}_\s  \rangle = \frac{\pi l}{2}\Delta  E \, ,
\eeq
so the FLEE in eq. (\ref{FLEE}) follows from 
 the saturation of the identity (\ref{relativeE-ineq}), i.e. 
 \beq
 \Delta \langle \mathcal{K}_\s \rangle  = \Delta S \, .
 \eeq
 
If we consider two nearby density matrices, i.e.
\beq
\s=\rho_0 \, , \qquad \rho=\rho_0 + \varepsilon \rho_1+ {O}(\varepsilon^2) \, ,
\label{nearby-density}
\eeq
where $\varepsilon$ is an expansion parameter, 
the relative entropy scales with $\varepsilon$ as follows
\beq
 S(\rho || \s) = {O}(\varepsilon^2) \, .
 \eeq 
Since the $ {O}(\varepsilon)$ contribution to relative entropy vanishes, 
there is a general expectation \cite{Blanco:2013joa} that
for small deformations eq. (\ref{relativeE-ineq}) is saturated.
However, the question if FLEE is satisfied in quantum field theory is subtle:
there the density matrix $\rho$ is infinite dimensional
and so it is not clear, in principle, when a perturbation might be considered small.

For a fixed value of the multitrace coupling $\kappa$, the monopole solution
can not be continuously deformed to the vacuum due to topological arguments.
For this reason, eq. (\ref{nearby-density}) can not be realised for 
arbitrary small $\varepsilon$. Then, the monopole deformation can not
be considered close enough to the vacuum and FLEE is not satisfied.


\section{Conclusions}
 \label{sect:conclusions}
 
 In this paper, we  studied a magnetic monopole solution in the
static global AdS$_4$ setup introduced in   \cite{Esposito:2017qpj}.
We found an approximate analytical solution for the monopole
in the regime of small gauge coupling $e$  which includes the leading-order backreaction on the metric.
By using the map introduced in \cite{Horowitz:1999gf},
the static monopole in global AdS$_4$ is mapped into a falling monopole in the Poincar\'e patch.
This bulk configuration is dual to a local quench on the CFT side.
The expectation values of local operators depend on the choice of boundary conditions.
With Dirichlet or Neumann conditions, the falling monopole is dual to a field theory
with a time-dependent source. With the special choice of multitrace deformation
in eq. (\ref{multitrace-speciale}), the monopole is dual to a field theory with zero sources.
In this case, there is no energy injection and the form of the energy-momentum  
tensor is the same as the one of a falling black hole  \cite{Nozaki:2013wia}.

Then, we computed the entanglement entropy due to the leading order 
classical bulk backreaction. We focused on a regime where
the monopole mass in eq. (\ref{massa-monopolo}) is sufficiently large
in order to neglect quantum corrections, see eqs.  (\ref{LGepsilon}).
The behaviour of entanglement entropy is  rather 
different compared to the case of the falling black hole.
For spherical regions centered on the local quench,
the perturbed entanglement entropy is always less than the vacuum value, i.e. $\Delta S \leq 0$.
This is consistent with the presence of a condensate
at the core of the local quench \cite{Albash:2012pd}.

In the case of a spherical region not centered at the origin,
there is a competition in $\Delta S$ between
the negative 
contribution from the condensate and the 
positive one due to quasiparticles \cite{Calabrese:2005in}.
Depending on the radius $l$ and on the distance $\xi$ from the origin of the spherical region,
 $\Delta S$ can be positive or negative, see figure \ref{trans_disk-plot}.
In the case of half-plane region,
the negative contribution to $\Delta S$ due to the condensate dominates
at early times, while the positive contribution due to quasiparticles
dominates at late times (see figure \ref{halfplane-plot}).

For a quench dual to a falling black hole,
the First Law of Entanglement Entropy (FLEE)  \cite{Bhattacharya:2012mi}
is satisfied for small subregions.
For the quench dual to the monopole, we found that the FLEE is not satisfied.
This is a feature shared with other AdS backgrounds which involve the backreaction
of scalar bulk fields, see \cite{Blanco:2013joa,OBannon:2016exv}.
On the field theory side, the violation of the FLEE 
comes from the non-saturation of the inequality in eq. (\ref{relativeE-ineq}).
It is still an open question whether a given deformation 
obeys or not the FLEE \cite{OBannon:2016exv}
in a quantum field theory. It would be interesting to further
investigate FLEE in non-equilibrium systems, in order to 
understand its general validity conditions.

Analytical  soliton solutions which include  backreaction 
 are quite rare in AdS spacetime. 
The monopole solution found in this paper 
can be the starting point for several further investigations.
In particular, it would be interesting to study 
more general solitonic objects in AdS.
For instance, vortex strings in AdS were considered by many authors
\cite{Albash:2009iq,Montull:2009fe,Keranen:2009re,Domenech:2010nf,Iqbal:2011bf,Dias:2013bwa,Maeda:2009vf,Tallarita:2019czh,Tallarita:2019amp}.
A static configuration in Poincar\'e patch with a monopole attached to a vortex 
string should also be possible, as proposed in \cite{Dias:2013bwa}:
the vortex string can pull the vortex and counterbalance
the gravitational force that makes it fall towards the center of AdS.
It would be interesting to find explicit solutions for these objects  and to investigate
their field theory duals.

Another possible direction is the study of holographic complexity \cite{Susskind:2014rva,Stanford:2014jda,Brown:2015bva}.
Quantum computational complexity is a recent quantum information entry in the holographic dictionary, 
which was motivated by the desire of understanding the growth of the Einstein-Rosen bridge
inside the event horizon of black holes.
Complexity for several examples of global and local quenches
has been studied by several authors, e.g. \cite{Moosa:2017yvt,Chapman:2018dem,Chapman:2018lsv,Chen:2018mcc,Auzzi:2019mah,DiGiulio:2021oal,Ageev:2018nye,Ageev:2019fxn,DiGiulio:2021noo}.
 It would be interesting to investigate complexity for quenches dual to a falling monopole.
 This analysis may give us useful  insights to understand the impact of condensates 
 on quantum complexity.

\section*{Acknowledgments}

We are grateful to Stefano Bolognesi for useful discussions.   
N.Z. acknowledges the Ermenegildo Zegna's Group for the financial support.

 \section*{Appendix}
\addtocontents{toc}{\protect\setcounter{tocdepth}{1}}
\appendix

\section{Equations of motion}

\subsection{Coordinate $r$}
\label{Appe-eqs-erre}

In the probe approximation, the equations of motion  are
\bea
F'' &=& - F' \frac{2 (1+2 r^2)}{r (1+r^2)}- F \frac{-2 + 2 r^2 + 3 e r F -e^2 r^2 F^2}{r^2 (1+r^2)}
-H^2 \frac{e (1- e r F)}{r(1+r^2)} \, ,
\nl
H''&=& -H' \frac{2 (1+ 2 r^2)}{r (1+r^2)}+2 H \, \frac{(1-e r F)^2}{r^2(1+r^2)}
-\frac{2 H }{1+r^2} \, .
\label{sistema-H-F-senza-backreaction}
\eea
Including backreaction, the full set of equations of motion is
\bea
F'' &=&
 -F' \frac{2  (1+ 2 r^2) }{ r (1+r^2)} 
- \frac{g' }{g  } \le \frac{F}{r } +F'\ri
-H^2 \frac{e (1- e r F)}{r (1+r^2)} \frac{h}{g} \nl
&& - F \frac{1}{{ r^2 (1+r^2)}}
\left( 2 r^2 
+\frac{h}{g} \left[ -2 +3 e r F - e^2 r^2 F^2 \right]
\right) \, ,
\nl 
H''&=& - H' \le \frac{2 (1+ 2 r^2)}{r (1+r^2)}
+\frac{g'}{g} \ri
- \frac{h}{g} \le
-2 H \, \frac{(1-e r F)^2}{r^2(1+r^2)}
+\frac{2 H }{1+r^2}  \ri  \, ,
\nl
g' &=&   \frac{4 \pi G}{L^2} h \, \frac{   
 2 H^2 \left[  (2-e r F) e r F +r^2-1 \right]
  -F^2 (2-e r F )^2  }{r (1+r^2)}
 + \frac{1+3 r^2}{r (r^2+1)} (h-g) 
   \, , \nl 
h' &=&\frac{8 \pi G}{L^2} h \le \frac{r}{2} (H')^2+ \frac{(F+r F')^2}{r} \ri \, .
\eea

\subsection{Coordinate $\psi$}
\label{Appe-eqs-psi}

In the probe approximation, we find
\bea
F'' &=& \frac{ e^2 F^3+F \left(e^2 H^2+2 \cot ^2(\psi )-2\right)-3 e
   F^2 \cot (\psi )-\cot (\psi ) \left(e H^2+2 F'\right) }{\cos^2 \psi}  \, ,
\nl
H'' &=& 4 \, \frac{  H \left(e^2 F^2 \tan (\psi )-2 e F+2 \cot (2 \psi)
   \right)-H' }{\sin(2 \psi ) }  \, .
   \label{no-backreaction-psi-vars}
\eea
Including backreaction, we get
\bea
 F''&=& \frac{1}{g}\left[-e \csc \psi \sec \psi \left(3 F^{2}+H^{2}\right)   h
 +e^{2} \sec^2 \psi \,  F^{3} h-F' \left(2 g \, \csc \psi \sec \psi +g^{\prime}\right)\right. \nl
& &  \left.+F\left(h  \left(2 \csc ^{2} \psi+e^{2} H^{2} \sec^{2} \psi \right)
-\sec \psi\left(2 g \, \sec \psi + g' \, \csc \psi \right)\right)\right] \, , \nl
 H'' &=& \frac{1}{g}\left[2 h H\left(\csc ^{2} \psi
-2 e F \, \csc \psi \sec \psi +\sec ^{2} \psi\left(-1+e^{2} F^{2} \right)\right)\right.\nl
&&\left.-\left(2 g \, \csc \psi \sec \psi +g' \right) H' \right]  \, , \nl
h^{\prime} &=& \frac{2 \pi G}{L^{2}} h 
\left[2\left( F \, \csc \psi \sec \psi  +F^{\prime}\right)^{2}+H^{\prime 2}\right]  \sin 2 \psi \, , \nl
g^{\prime} &=& \tan \psi\left(3+\cot ^{2} \psi\right)(h-g)+\frac{4 \pi G}{L^{2}} h \tan \psi\left[-F^{2}(-2 \cot \psi+e F)^{2}\right. 
\nl
& & \left. + 2 H^{2}-2(\cot \psi-e F)^{2} H^{2} \right] \, .
\eea


\section{Abelian field strength and flux}
\label{flux-appendix}

The abelian field strength and its dual are   \cite{tHooft:1974kcl}:
\beq
 \mathcal{F}_{\mu \nu} = n^a F^a_{\mu \nu} -\frac{1}{e}\epsilon^{a b c} n^a D_\mu n^b D_\nu n^c \, ,
 \qquad
 \tilde{\mathcal{F}}^{\mu \nu} \equiv \frac12 \frac{\epsilon^{\mu \nu \a \b}}{\sqrt{-g}}  \mathcal{F}_{\a \b} \, ,
\eeq
which satisfies 
\beq
D_\mu \tilde{\mathcal{F}}^{\mu \nu} =\frac{4 \pi}{e} k^\nu \, ,
\qquad
k_\mu=\frac{1}{8 \pi} \epsilon_{\mu \nu \rho \s} \epsilon_{abc} \p^\nu n^a   \p^\rho n^b   \p^\s n^c \, ,
\eeq
where $k_\mu$ is the topological current. 
The only non-vanishing components of the dual electromagnetic tensor $\tilde{\mathcal{F}}^{\mu \nu}$ are:
\beq
\tilde{\mathcal{F}}^{t r}= -\tilde{\mathcal{F}}^{r t}= -\frac{1}{e r^2}\, .
\eeq
The magnetic flux on a sphere of radius $r$ is given by Stokes theorem:
\beq
Q=\int_{S^2} \tilde{\mathcal{F}}^{\mu \nu} d S_{\mu \nu} \, , \qquad
d S_{\mu \nu}=n_{[ \mu } r_{\nu]} \, r^2 \sin \theta \, d \theta d \varphi \, ,
\eeq
where $n_{\mu}$ and $r_{\nu}$ are the unit time and radial vectors, respectively. 
A direct computation gives:
\beq
Q=-\frac{8 \pi }{e} \, .
\eeq
Note that the magnetic flux is topological and  independent of the profile functions details.


\section{Details of the boundary energy-momentum tensor}
 \label{appe-energy-momentum-tensor}

\subsection{Black hole }
 \label{appe-energy-momentum-tensor:BH}

In Fefferman-Graham (FG) coordinates, the metric has the following form
\beq
\hhx^\mu=(\hhz,\hht,\hhx,\hhphi) \, 
\qquad
ds^2=L^2 \le
\frac{d \hhz^2}{\hhz^2}+ \frac{1}{\hhz^2} g_{ab}(\hhz,\hhx^a) d \hhx^a d \hhx^b
 \ri \, ,
\eeq
where the index $a$ runs on boundary coordinates
\beq
\hhx^a=(\hht,\hhx,\hhphi) \, ,
\eeq
and we take $\hhphi=\vphi$.
The FG coordinates can be built in a perturbative way nearby the boundary, i.e.
\beq
z= \hhz + \sum_{k=2}^{\infty} a_k( \hhx, \hht) \hhz^k \, , \qquad
x= \hhx + \sum_{k=1}^{\infty} b_k( \hhx, \hht) \hhz^k \, , \qquad
t= \hht + \sum_{k=1}^{\infty} c_k( \hhx, \hht) \hhz^k \, .
\label{FG-expansion}
\eeq
Plugging into the metric in Poincar\'e coordinates and comparing with the FG metric order by order, we get:
\bea
b_1&=& c_1=0 \, , \qquad a_2=b_2=c_2=0 \, , \qquad
a_3=b_3=c_3=0 \, \qquad b_4=c_4=0 \, ,
\nl
a_4&=& -\frac{4 \, A^3 M}
{3 \left( A^4 + 2 A^2 \hht^2 + 2 A^2 \hhx^2+\hht^4-2 \hht^2
   \hhx^2+\hhx^4\right)^{3/2}} \, .
\eea

The energy-momentum tensor can be obtained from the results of \cite{Balasubramanian:1999re} 
\beq
T_{mn}^{(BH)}=\frac{L}{8 \pi G } \,  \lim_{\hhz \rightarrow 0}
\frac{1}{\hhz} \le K_{mn} -\gamma_{mn} K -\frac{2}{L} \gamma_{mn}    \ri \, .
\label{T-boundary-blackhole}
\eeq
Here $\gamma_{mn}$ is the induced metric on a $\hhz$-constant surface nearby the boundary,
$K_{mn}$ denotes the extrinsic curvature tensor calculated with 
an inward unit vector normal to the $\hhz$-constant
surface, and $K$ is the trace of the extrinsic curvature tensor.

To explicitly write the components of the energy-momentum tensor, it is convenient to introduce the lightcone coordinates: 
 \beq
 u^m=(u,v,\varphi) \, , \qquad u= t-x \, , \qquad v= t +x \, .
 \eeq
 In these coordinates, the non-vanishing elements of $T_{mn}^{(BH)}$ are
\bea
\label{T-cono-luce-BH}
T_{uu}^{(BH)} &=& \frac{A^3 L^2  M }{8 \pi  \, G } 
 \frac{3 }{ \left(A^2+u^2\right)^{5/2} { \left(A^2+v^2\right)}^{1/2}} \, , 
\nl
T_{vv}^{(BH)} &=&   \frac{A^3 L^2  M }{8 \pi  \, G } 
\frac{3 }{ \left(A^2+v^2\right)^{5/2} { \left(A^2+u^2\right)}^{1/2}} \, ,
\nl
T_{uv}^{(BH)} &=&  \frac{A^3 L^2  M }{8 \pi  \, G } 
\frac{1}{ \left(A^2+u^2\right)^{3/2} \left(A^2+v^2\right)^{3/2}} \, ,
\nl
T_{\varphi \varphi}^{(BH)} &=&  \frac{A^3 L^2  M }{8 \pi  \, G } 
 \frac{ (u-v)^2}{ \left(A^2+u^2\right)^{3/2}  \left(A^2+v^2\right)^{3/2}} \, .
\eea


\subsection{Monopole with Dirichlet boundary conditions}
 \label{appe-energy-momentum-tensor:D}

In order to put the metric with monopole backreaction in FG coordinates,
 we consider the expansion of $h$ and $g$ nearby the boundary
 in eq. (\ref{hg-expansion}), setting also $h_2=-g_2$. 
Then, using the change of variables in eq. (\ref{FG-expansion})
and solving order by order, we obtain
\bea
b_1&=&c_1=0 \, , \qquad a_2=b_2=c_2=0 \, , \qquad
b_3=c_3=0 \, ,
\nl
a_3&=& \frac{2 A^2 \, g_2 }{A^4+2 A^2 \hht^2+2 A^2 \hhx^2+\hht^4-2 \hht^2 \hhx^2+\hhx^4} \, , \nl
a_4&=& \frac{4 A^3 \left(g_3-h_3\right)}
{3 \left( A^4+2 A^2 \hht^2+2 A^2 \hhx^2+\hht^4-2
   \hht^2 \hhx^2+\hhx^4\right)^{3/2}}\, , \nl
b_4&=& -\frac{2 A^2 \, g_2  \hhx \left( A^2-\hht^2+\hhx^2\right)}
{\left( A^4+2 A^2 \hht^2+2 A^2 \hhx^2+\hht^4-2 \hht^2 \hhx^2+\hhx^4\right)^2} \, , \nl
c_4&=& \frac{2 A^2 \, g_2 \hht \left( A^2+\hht^2-\hhx^2\right)}
{\left( A^4+2 A^2 \hht^2 + 2 A^2 \hhx^2+\hht^4-2 \hht^2 \hhx^2+\hhx^4\right)^2} \, . 
\eea
We can now use the generalisation of eq. (\ref{T-boundary-blackhole}) involving scalars 
 \cite{deHaro:2000vlm} to extract the energy-momentum tensor
\beq
T_{mn}^{(D)}=\frac{L}{8 \pi G } \,  \lim_{\hhz \rightarrow 0}
\frac{1}{\hhz} \le K_{mn} -\gamma_{mn} K -\frac{2}{L} \gamma_{mn} 
-4 \pi G \frac{\gamma_{mn}}{L}   \phi^a \phi^a    \ri \, .
\eeq
The elements of the energy-momentum tensor in lightcone coordinates
look qualitatively similar to the corresponding elements computed in the BH background,
see eq. (\ref{T-cono-luce-BH}):
\bea
T_{uu}^{(D)} &=& T_{uu}^{(BH)} \le \frac{16 \pi G \a_H \b_H-3 L^2 g_3}{3 L^2 M} \ri \, , 
\nl
T_{vv}^{(D)}&=&T_{vv}^{(BH)} \le \frac{16 \pi G \a_H \b_H-3 L^2 g_3}{3 L^2 M} \ri \, ,
\nl
T_{uv}^{(D)} &=& T_{uv}^{(BH)} \le \frac{-16 \pi G \a_H \b_H-3 L^2 g_3}{3 L^2 M} \ri \, , 
\nl 
T_{\varphi \varphi}^{(D)}&=&T_{\varphi \varphi}^{(BH)} \le \frac{32 \pi G \a_H \b_H-3 L^2 g_3}{3 L^2 M} \ri \, .
\label{T-cono-luce-D}
\eea
The Ward identity for  $T_{mn}^{(D)}$ gives
\beq
\p^m T_{mn}^{(D)}= \tilde{\b}_H \p_n \tilde{\a}_H
= \langle \mathcal{O}_2 \rangle \p_n J_D \, ,
\eeq
and the trace of the energy-momentum tensor is
\beq
\eta^{mn} T_{mn}^{(D)}= \ta_H \tb_H= \langle \mathcal{O}_2 \rangle  J_D \, .
\eeq


\subsection{Monopole with Neumann and multitrace boundary conditions}
 \label{appe-energy-momentum-tensor:multitrace}

We will follow the approach in \cite{Caldarelli:2016nni} to determine
 the boundary energy-momentum tensor for multitrace deformations.
 These boundary conditions correspond to adding to the renormalised action $S_{\rm ren}$
a finite boundary action $S_{\mathcal{F}}$ given by 
\beq
S_{\mathcal{F}}= \int d^3 x \, \sqrt{-g_0} \, (J_F \ta_H+\mathcal{F}(\ta_H)) \, ,
\qquad J_{\mathcal{F}}=-\tilde{\b}_H - \mathcal{F}'(\tilde{\a}_H) \, ,
\eeq
where $\sqrt{-g_0}$ is the determinant of the boundary metric (at the end of the calculation
we will specialise to the Minkowski metric).
The variations of the action functionals are:
\bea
\delta S_{\rm ren}&=&  \int d^3 x \, \sqrt{-g_0} \le \frac{1}{2} T^{ij} (\delta g_{(0)})_{ij} +\tb_H \delta \ta_H  \ri \, ,
\nl
\d S_{\mathcal{F}}&=& \int d^3 x \, \sqrt{-g_0} \le 
- \ta_H \delta \tb_H -\ta_H \mathcal{F}''(\ta_H) \delta \ta_H - \tb_H \delta \ta_H
\ri  \, ,
\eea
so the total variation is
\bea
\delta S&=&\delta S_{\rm ren}+ \delta S_{\mathcal{F}} =
 \int d^3 x \sqrt{-g_0}  \le \frac{1}{2} T^{ij} (\delta g_{(0)})_{ij}  
 - \ta_H \delta \tb_H -\ta_H \mathcal{F}''(\ta_H) \delta \ta_H \ri \nl
 &=&  \int d^3 x \sqrt{-g_0}  \le \frac{1}{2} T^{ij} (\delta g_{(0)})_{ij}  
 + \ta_H \delta J_{\mathcal{F}} \ri \, .
\eea

Due to the shift of the action, there is also a shift in the energy-momentum tensor:
\beq
T^{(\mathcal{F})}_{ij}= T_{ij}^{(D)}+\eta_{ij} [ \mathcal{F}(\ta_H) +\ta_H J_F ]=
T_{ij}^{(D)}+\eta_{ij} [\mathcal{F}(\ta_H) -\ta_H \tb_H - \mathcal{F}'(\ta_H) \ta_H ] \, .
\eeq
The divergence of the energy-momentum tensor is:
\bea
\p^i T^{(\mathcal{F})}_{ij}&=&\p^i T_{ij}^{(D)} + \p_j  [\mathcal{F}(\ta_H) -\ta_H \tb_H - \mathcal{F}'(\ta_H) \ta_H ]
\nl
&=& \tb_H \p_j \ta_H + \p_j  [\mathcal{F}(\ta_H) -\ta_H \tb_H - \mathcal{F}'(\ta_H) \ta_H ] \nl
&=& -\ta_H (\p_j \tb_H +\mathcal{F}'' (\ta_H) \,  \p_j \ta_H) \, ,
\eea
while the trace is:
\bea
\eta^{ij} T^{(\mathcal{F})}_{ij} &=& (T^{(D)})^i_i + 3  [\mathcal{F}(\ta_H) -\ta_H \tb_H - \mathcal{F}'(\ta_H) \ta_H ] \nl
&=& -2 \ta_H \tb_H +3 \mathcal{F}(\ta_H) -3 \ta_H \mathcal{F}'(\ta_H) \, .
\eea
Setting to zero the source $J_{\mathcal{F}}$ corresponds to
\beq
\tb_H=-\mathcal{F}'(\ta_H ) \, .
\eeq
Note that in this case $T^{(\mathcal{F})}_{ij}$ is conserved, i.e. $\p^i T^{(\mathcal{F})}_{ij}=0$.

\section{Details of the entanglement entropy calculations}

\subsection{Translated-disk region}
\label{appe:entanglement-entropy-trans-disk}

Defining the polar-like coordinates
\beq
x_1 = \xi + p \, \cos \vartheta \, , \qquad x_2 =  p \, \sin \vartheta \, ,
\eeq
the entanglement entropy is given by
\beq
\begin{aligned}
\Delta S & = \frac{\pi \alpha_H^2}{2 l} \int_0^l dp \frac{p}{\le l^2 - p^2 \ri^{3/2}} \int_0^{2 \pi} d\vartheta \left[ \frac{\le h_{\varepsilon} + g_{\varepsilon} \ri \le p^2 \cos^2 \vartheta - l^2 \ri \xi^2 t^2}{\omega_{\xi}} \right. \\
& \left. + \frac{\le h_{\varepsilon} - g_{\varepsilon} \ri}{4} \frac{\left[ p \, \omega_{\xi} + 2 \xi \cos \vartheta \le l^2 - p^2 \ri \le l^2 -t^2 +A^2 +\xi^2 + 2 \xi p \cos \vartheta \ri \right]^2}{\omega_{\xi} \left[ \omega_{\xi} + 4 A^2 \le p^2 - l^2 \ri \right]} \right. \\
& \left. + \frac{\le h_{\varepsilon} - g_{\varepsilon} \ri l^2 \xi^2 \sin^2 \vartheta \le l^2 - p^2 \ri \le l^2 -t^2 +A^2 +\xi^2 + 2 \xi p \cos \vartheta \ri^2}{\omega_{\xi} \left[ \omega_{\xi} + 4 A^2 \le p^2 - l^2 \ri \right]} \right] \, ,
\end{aligned}
\eeq
where
\beq
\omega_{\xi} \equiv \omega ( \sqrt{l^2 + \xi^2 + 2 \, \xi \, p \cos \le 2 \vartheta \ri},t) \, ,
\eeq
and $\omega$ is defined in eq. (\ref{omega}). In the above integral, both $h_{\varepsilon}$ and $g_{\varepsilon}$ are functions of
\beq
r = \frac{1}{2 A} \sqrt{\frac{\omega_{\xi} + 4 A^2 \le p^2 - l^2 \ri}{l^2 - p^2}} \, .
\eeq

\subsection{Half-plane region}
\label{appe:entanglement-entropy-half-plane}
 
 As in eq. (\ref{rescale-1}),  the $A$ dependence can be completely reabsorbed
introducing the rescaled quantities  
\beq
z \to \frac{z}{A} \, , \qquad  x_2 \to \frac{x_2}{A} \, , \qquad  t \to \frac{t}{A} \, .
\eeq
Consequently, without loss of generality,  from now on we set $A=1$.

For general $t>0$, we can write a closed-form expression for the entropy variation
\beq
\label{integrale-entropia-semipiano}
\begin{aligned}
\Delta S & = &  \frac{\pi \a_H^2}{8} \int_0^{+ \infty} dz \int_{- \infty}^{+ \infty} dx_2 \left[  \frac{\le h_{\varepsilon} - g_{\varepsilon} \ri \le C^2 + 4 z^2 x_2^2 \, D^2 \ri}{z^2 \le D^2 + 4  \, t^2 \ri^2} - 4 t^2 \frac{\le h_{\varepsilon} + g_{\varepsilon} \ri \le x_2^2 + z^2 \ri}{z^2 \le D^2 + 4  \, t^2 \ri} \right] \, ,
\end{aligned}
\eeq
with
\beq
C \equiv  t^4 -z^4 -2 t^2 \le x_2^2 -1 \ri + \le x_2^2 + 1 \ri^2 \, , \qquad D \equiv -t^2 +x_2^2 +z^2 +1 \, .
\eeq
In the above expression, $h_{\varepsilon}(r)$ and $g_{\varepsilon}(r)$ are the functions 
defined in  eq. (\ref{definizione-h-g-epsilon}), with
\beq
r = \ \frac{ \sqrt{\le -t^2 +x_2^2 +z^2 \ri^2 + 2 \le t^2 +x_2^2 -z^2 \ri + 1}}{2  \, z} \, .
\eeq
In order to understand the large $t$ behaviour of the entropy, 
it is convenient to introduce the variables $\rho, \gamma$
\beq
z= \rho \cos \gamma \, , 
\qquad 
x_2 = \rho \sin \gamma \, .
\eeq	
At large $t$, the integrand in eq. (\ref{integrale-entropia-semipiano})
is non-vanishing just in a region at $\rho=t \pm \mu $, with $\mu$ of order $1$.
For convenience, we can introduce
\beq
\rho= t + \delta \, .
\eeq
It turns out that, at large $t$,
the term proportional to $(h_{\varepsilon} - g_{\varepsilon})$ in eq. (\ref{integrale-entropia-semipiano}) is much smaller
than the one proportional to $(h_{\varepsilon} + g_{\varepsilon})$.
Moreover, at large $t$, we can use the approximation
\beq
r = \frac{\sqrt{\delta^2 + \sin^2 \gamma} }{\cos \gamma} + O(1/t) \, .
\eeq
At the leading order in $t$, we find
\beq
\Delta S = K \, \a_H^2 \,  t  \,  ,
\eeq
where
\beq
K=-\frac{\pi}{2} \, \int_0^{\infty} \frac{r}{1+r^2} \mathbb{E}(-r^2) [h_\ep(r) + g_\ep(r)] \, dr \, .
\label{K-expression}
\eeq
$\mathbb{E}$ is the complete elliptic integral, defined as follows
\beq
\mathbb{E}(m)=\int_0^{\pi/2} d \gamma \, \sqrt{1-m \sin^2 \gamma} \, .
\eeq 
Numerically, we get the approximate value $K \approx 0.636$.


\bibliography{at}
\bibliographystyle{at}

\end{document}